\documentclass[prb,aps,amsmath,amssymb,superscriptaddress,floatfix]{revtex4}
\usepackage{graphicx}
\usepackage[american]{babel}

\usepackage{pspicture}
\usepackage{pstricks}

\begin{document}
\newcommand{\hexa}{ \; \psset{unit=0.7cm} \pspicture[0.2](1,0.5) \psdots[linecolor=black,dotsize=.15](0.25,0)(0.25,0.5) (1,0.25)  \psset{linewidth=0.03,linestyle=solid} \pspolygon[](0,0.25)(0.25,0)(0.75,0)(1,0.25) (0.75, 0.5) (0.25, 0.5) \psset{linewidth=0.03,linestyle=solid} \pspolygon[fillcolor=lightgray, fillstyle=solid](0.25,0.0)(0.75,0)(0.5,0.25) \pspolygon[fillcolor=lightgray, fillstyle=solid](0.25,0.5)(0.75,0.5)(0.5,0.25) \psline[](0.5,0)(0.5,0.5)  \psset{linewidth=0.08,linestyle=solid} \endpspicture\;}

\newcommand{\hexb}{ \; \psset{unit=0.7cm} \pspicture[0.2](1,0.5)  \psdots[linecolor=black,dotsize=.15](0,0.25)(0.75,0) (0.75,0.5)  \psset{linewidth=0.03,linestyle=solid} \pspolygon[](0,0.25)(0.25,0)(0.75,0)(1,0.25) (0.75, 0.5) (0.25, 0.5) \psset{linewidth=0.03,linestyle=solid} \pspolygon[fillcolor=lightgray, fillstyle=solid](0.25,0.0)(0.75,0)(0.5,0.25) \pspolygon[fillcolor=lightgray, fillstyle=solid](0.25,0.5)(0.75,0.5)(0.5,0.25) \psline[](0.5,0)(0.5,0.5)

\psset{linewidth=0.08,linestyle=solid} \endpspicture\;}

\newcommand{\hexaf}{ \; \psset{unit=0.7cm} \pspicture[0.2](1,0.5) \psset{linewidth=0.03,linestyle=solid} \pspolygon[](0,0.25)(0.25,0)(0.75,0)(1,0.25) (0.75, 0.5) (0.25, 0.5) \psset{linewidth=0.03,linestyle=solid} \pspolygon[fillcolor=lightgray, fillstyle=solid](0.25,0.0)(0.75,0)(0.5,0.25) \pspolygon[fillcolor=lightgray, fillstyle=solid](0.25,0.5)(0.75,0.5)(0.5,0.25) \psline[](0.5,0)(0.5,0.5)  \psset{linewidth=0.08,linestyle=solid} \psdots[linecolor=black,dotsize=.15](0.25,0)(0.25,0.5) (1,0.25)(0.5,0.25)  \endpspicture\;}

\newcommand{\hexbf}{ \; \psset{unit=0.7cm} \pspicture[0.2](1,0.5) \psset{linewidth=0.03,linestyle=solid} \pspolygon[](0,0.25)(0.25,0)(0.75,0)(1,0.25) (0.75, 0.5) (0.25, 0.5) \psset{linewidth=0.03,linestyle=solid} \pspolygon[fillcolor=lightgray, fillstyle=solid](0.25,0.0)(0.75,0)(0.5,0.25) \pspolygon[fillcolor=lightgray, fillstyle=solid](0.25,0.5)(0.75,0.5)(0.5,0.25) \psline[](0.5,0)(0.5,0.5)  \psset{linewidth=0.08,linestyle=solid} \psdots[linecolor=black,dotsize=.15](0,0.25)(0.75,0) (0.75,0.5)(0.5,0.25) \endpspicture\;}

\newcommand{\hexafg}{ \; \psset{unit=0.7cm} \pspicture[0.2](1,0.5)  \psset{linewidth=0.03,linestyle=solid} \pspolygon[](0,0.25)(0.25,0)(0.75,0)(1,0.25) (0.75, 0.5) (0.25, 0.5) \psset{linewidth=0.03,linestyle=solid} \pspolygon[fillcolor=lightgray, fillstyle=solid](0.25,0.0)(0.75,0)(0.5,0.25) \pspolygon[fillcolor=lightgray, fillstyle=solid](0.25,0.5)(0.75,0.5)(0.5,0.25) \psline[](0.5,0)(0.5,0.5)  \psset{linewidth=0.08,linestyle=solid} \psdots[linecolor=black,dotsize=.15](0.25,0)(0.25,0.5)(1,0.25) \psdots[linecolor=gray,dotsize=.15](0.5,0.25)  \endpspicture\;}

\newcommand{\hexbfg}{ \; \psset{unit=0.7cm} \pspicture[0.2](1,0.5) \psset{linewidth=0.03,linestyle=solid} \pspolygon[](0,0.25)(0.25,0)(0.75,0)(1,0.25) (0.75, 0.5) (0.25, 0.5) \psset{linewidth=0.03,linestyle=solid} \pspolygon[fillcolor=lightgray, fillstyle=solid](0.25,0.0)(0.75,0)(0.5,0.25) \pspolygon[fillcolor=lightgray, fillstyle=solid](0.25,0.5)(0.75,0.5)(0.5,0.25) \psline[](0.5,0)(0.5,0.5)  \psset{linewidth=0.08,linestyle=solid} \psdots[linecolor=black,dotsize=.15](0,0.25)(0.75,0) (0.75,0.5) \psdots[linecolor=gray,dotsize=.15](0.5,0.25)  \endpspicture\;}

\newcommand{\smallhexv}{\; \psset{unit=0.7cm} \pspicture[0.15](0.2,0.35) \psset{linewidth=0.03,linestyle=solid}    \pspolygon[](0.1,-0.1)(0,0.0)(0,0.2)(0.1,0.3)(0.2,0.20)(0.2,-0.0)  \endpspicture \;}

\newcommand{\smallhexh}{ \psset{unit=0.7cm} \pspicture[0.2](0.4,0.2) \psset{linewidth=0.03,linestyle=solid}    \pspolygon[](0,0.1)(0.1,0.0)(0.3,0)(0.4,0.1)(0.3,0.2)(0.1,0.2)  \endpspicture\;}

\title{FRACTIONALLY CHARGED EXCITATIONS ON FRUSTRATED LATTICES}

\author{E. Runge }

\address{Technische Universit\"at Ilmemau, Fakult\"at f\"ur Mathematik und Naturwissenschaften,
FG Theoretische Physik I, 98693 Ilmenau, Germany}

\author{F. Pollmann and P. Fulde}

\address{Max-Planck-Institut f\"ur Physik komplexer Systeme, N\"othnitzer Stra\ss e 38, 01187 Dresden, Germany}

\begin{abstract}
Systems of strongly correlated fermions on certain geometrically frustrated lattices at particular filling factors support excitations with fractional charges $\pm e/2$. We calculate quantum mechanical ground states, low--lying excitations and spectral functions of finite lattices by means of numerical diagonalization. The ground state of the most thoroughfully studied case, the criss-crossed checkerboard lattice, is degenerate and shows long--range order.  Static fractional charges are confined by a weak linear force, most probably leading to bound states of large spatial extent. Consequently, the quasi-particle weight is reduced, which reflects the internal dynamics of the fractionally charged excitations. By using an additional parameter, we fine--tune the system to a special point at which fractional charges are manifestly deconfined---the so--called Rokhsar--Kivelson point. For a deeper understanding of the low--energy physics of these models and for numerical advantages,  several conserved quantum numbers are identified. 
\end{abstract}

\keywords{fractional charge, strongly correlated particles, confinement}

\maketitle

\section{Introduction}

Quantization of charge is a very basic feature in the description
of the physical world. Therefore, fractionally charged excitations
came as a surprise to physicists. Already back in the year 1979, Su,
Schrieffer, and Heeger \cite{su1979} showed that a model describing
the one-dimensional (1D) chain molecule trans--polyacetylene $\mathrm{(CH)_{n}}$
supports excitations with spin--charge separation. This is not yet
charge fractionalization, but when the model is considered at different
electron densities (corresponding to extremely high doping), it turns
out that at certain filling factors elementary excitations with fractional
charge exist {[}in the simplest case $q=\pm e/3\ $and $q=\pm2e/3${]}.
A few years later, Laughlin \cite{laughlin1983} interpreted the much
celebrated fractional quantum Hall effect (FQHE) in terms of fractionally
charged (quasi-) particles (fcp) and fractionally charged (quasi-)
holes (fch). Thereby, he firmly established the idea of fractional
charges in our understanding of solid--state physics. Direct electron--electron
interactions are crucial in the FQHE: Correlations become strong in
an applied magnetic field because the kinetic energy of the electrons
is quenched. Of course, no one will ever extract a fraction of an
electron from an FQHE sample. But thinking of an extra electron or
an extra hole near fill factor $\nu=1/3$ as decaying into three separate
entities of charge $q=\pm e/3\ $each proved enormously helpful for
a qualitative and quantitative understanding of the FQHE---and was
rewarded by the Nobel Prize in Physics 1998. 

The question was left open whether or not fractionally charged excitations
exist in 2D or 3D systems without a magnetic field. In 2002, it was
suggested by one of us \cite{fulde2002} that excitations with charge
$\pm e/2$ do exist in certain 2D and 3D lattices, e.g., the pyrochlore
lattice, which is a prototype 3D structure with geometrical frustration.
The original work was motivated by the transition metal compound $\mathrm{LiV_{2}O_{4}}$,
which surprisingly shows heavy--fermion behavior with, e.g., a large
$\gamma$~coefficient in the low-temperature specific heat $C=\gamma T$.\cite{kondo1997}
However, we will not address the issue whether the models discussed
here apply to any particular material or artifical systems such as
optical lattices, see e.g. Ref. \cite{hofstetter2002} and citations therein.
Instead, we try to contribute to the very general question whether
or not fractional charges can exist at all in truly 2D or 3D systems
in the absence of magnetic fields. 

We would like to argue that general prerequisites for fractional charges
are---as in the FQHE case---strong short--range correlations and
certain band fillings. Furthermore, the short--range correlations should
be somehow incompatible with the lattice structure in order to prevent
the development of long--range order. Following the general usage,
we will simply call the lattice {}``frustrated'' even though calling
the interaction {}``frustrated by the lattice'' would be the more
accurate terminology. 

In this contribution, we study the charge degrees of freedom on such
lattices systematically and consider a class of models of strongly
correlated spinless fermions. Most of our numerical calculations were
done for the 2D checkerboard lattice, which is easier to deal with
than the even more interesting 3D cases. For future work, one can
hope to learn from comparison with spin systems where numerous studies
exist for the pyrochlore lattice and its 2D relatives, the crisscrossed
checkerboard lattice and the kagome lattice.\cite{misguich2003,diep2005,hermele2004,lauchli2004,shannon2004}

\section{Fractional charges on frustrated lattices\label{sec:Fractional-charges-on-frutrated}}

\begin{figure}
\begin{center}\begin{tabular}{ccccc}
(a)~\includegraphics[%
  height=25mm,
  keepaspectratio]{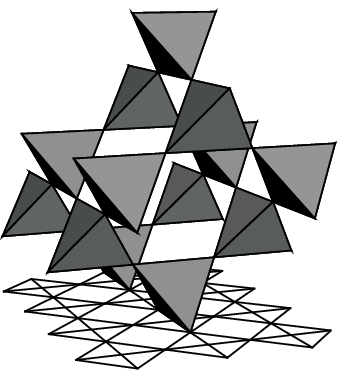}&
&
(b)~\includegraphics[%
  height=25mm,
  keepaspectratio]{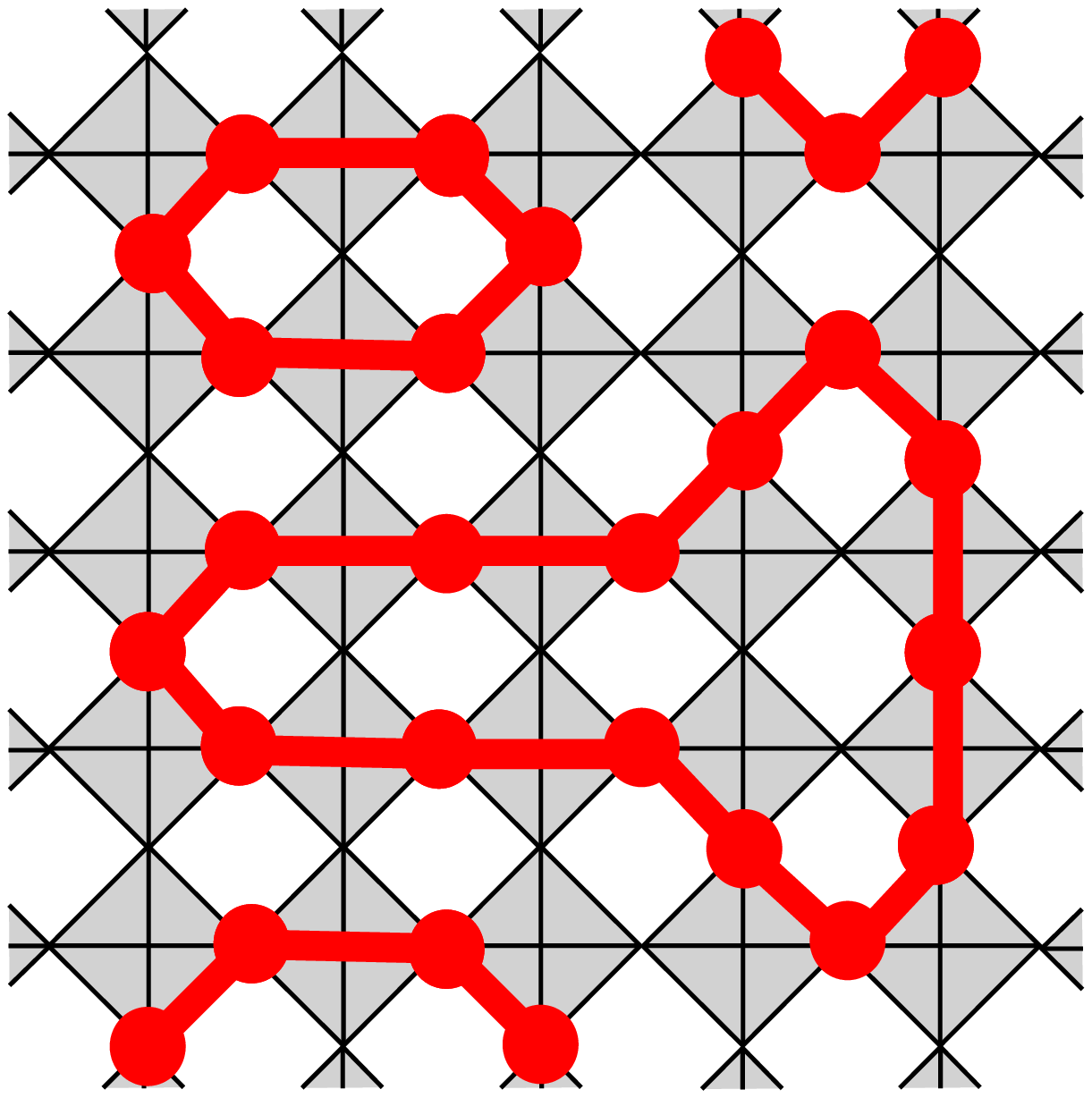}&
&
(c)~\includegraphics[%
  clip,
  height=25mm,
  keepaspectratio]{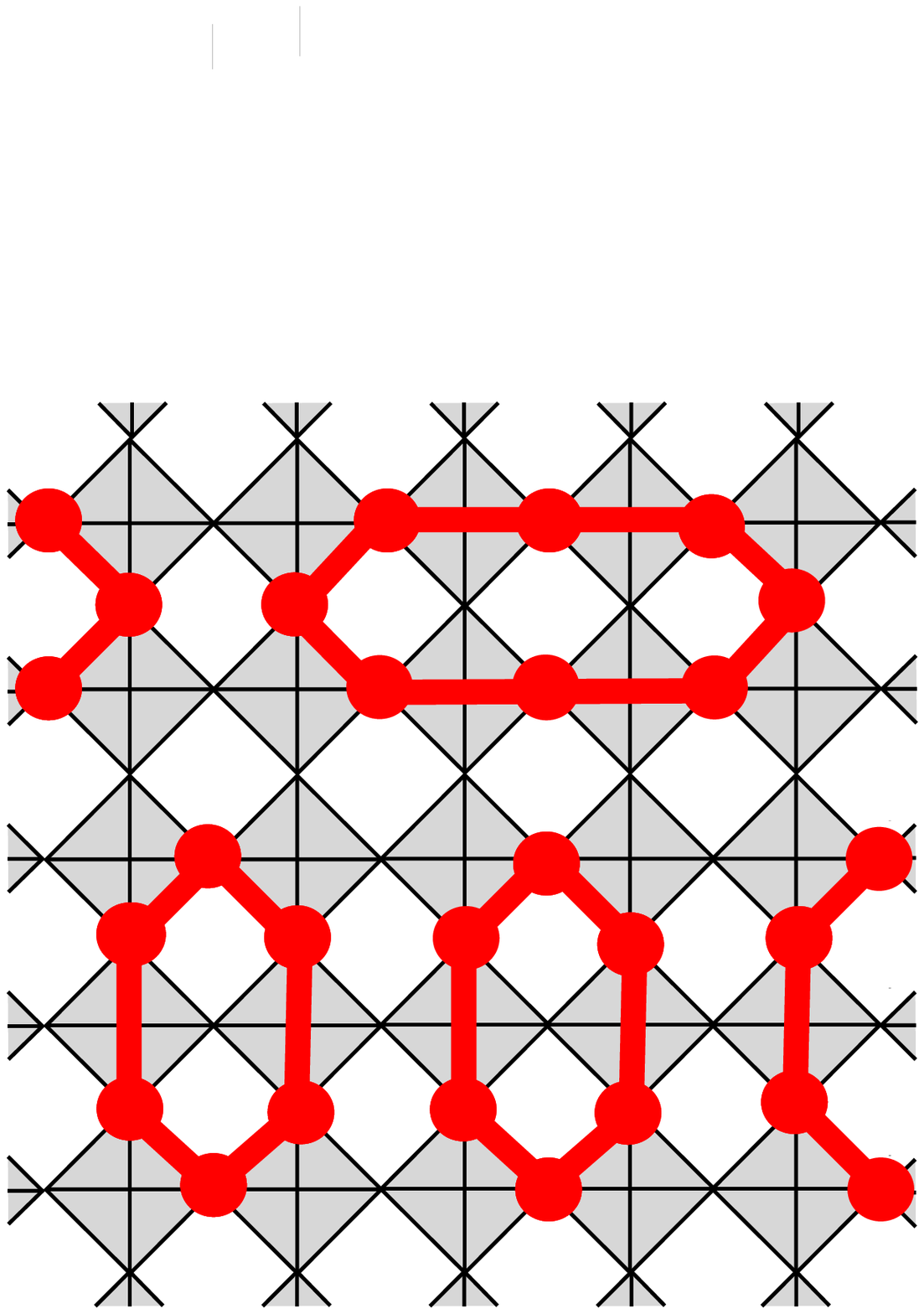}
\end{tabular}\end{center}
\caption{(a) Checkerboard lattice as 2D projection of the 3D pyrochlore lattice.
\cite{moessner2004} \label{cap:C2_pyro_checker} (b,c) Two examples
of allowed configurations on a checkerboard lattice at half filling.
Occupied sites are connected by thick lines as guides to the eye.
\label{cap:C2_allowed}}
\end{figure}

In order to illustrate the concept of fractional charges on frustrated
lattices, we focus here on the (crisscrossed) 2D checkerboard lattice.
Figure~\ref{cap:C2_pyro_checker}(a) illustrates that the checkerboard
lattice can be thought of as a projection of the pyrochlore lattice
onto a plane. In the following, we adopt the ideas of Ref.~\cite{fulde2002}
and consider a model Hamiltonian of spinless fermions\begin{equation}
H=-t\sum_{\langle i,j\rangle}\left(c_{i}^{\dag}c_{j}^{\vphantom{\dag}}+\mbox{H.c.}\right)+V\sum_{\langle i,j\rangle}n_{i}n_{j}\label{eq:C2_spinless_hamil}\end{equation}
on a checkerboard lattice. The operators $c_{i}^{\vphantom{\dag}}(c_{i}^{\dag})$
annihilate (create) fermions on sites $i$. The density operators
are $n_{i}=c_{i}^{\dag}c_{i}^{\vphantom{\dag}}$. We assume half filling,
$\sum_{i}n_{i}=\frac{N}{2}$ for a system with $N$ sites. Our main
interest is the regime where the nearest--neighbor hopping $t$ 
is much smaller than the nearest--neighbor repulsion $V$, i.e., $|t|/V\ll1$. 
Henceforth, we assume $t>0$. In analogy to the tetrahedra in a pyrochlore lattice, all bonds in a crossed square are equivalent.

\paragraph{Classical correlations and ground--state degeneracy.}

For a moment, let us set the hopping--matrix element $t$ to zero. The ground--state manifold is then \emph{macroscopically degenerate}: Every configuration that satisfies the so--called tetrahedron rule of having \emph{exactly} two particles on \emph{each} tetrahedron (crisscrossed square) is
a (classical) ground state \cite{anderson1956} and will henceforth be referred
to as an {}``allowed configuration.'' Examples are shown in Fig.~\ref{cap:C2_allowed}(b)
and (c). Note that our coordinate system is rotated by $45^{\circ}$ relative
to that of, e.g., Ref.~\cite{runge2004}. The resulting
difference in boundary conditions can lead to noticeable numerical
differences in particular for small cluster sizes. 

One can visualize the origin of the macroscopic degeneracy as follows:
Take the set of the six allowed crisscrossed squares and construct
row by row a larger allowed configuration. Whenever a crisscrossed
square is added, we can choose between one, two, or three different possibilities,
depending on the neighboring crisscrossed squares. Since there often
\emph{is} a choice to make, an exponential number of different allowed configurations 
can be constructed. Thus, the system has a finite  entropy at $T=0$. 
The exact value of the ground-state
degeneracy can be obtained from a mapping to the so--called six--vertex
model. \cite{baxter1982} Its solution is highly non--trivial due to
the existence of long--range correlations, which seem to be a generic
feature seen in many frustrated lattice models.\cite{pollmann2006a}

All classical ($t=0$) ground states have the important property of
being incompressible in the sense that no fermion can hop to another
empty site without creating defects and thereby increasing the repulsion energy.
In other words, we have to violate the tetrahedron rule in intermediate
states if we want to connect via hopping processes one allowed configuration
with another.

\paragraph{Fractional charges.}

\begin{figure}
\begin{center}\begin{tabular}{ccc}
(a)~\includegraphics[%
  height=22mm,
  keepaspectratio]{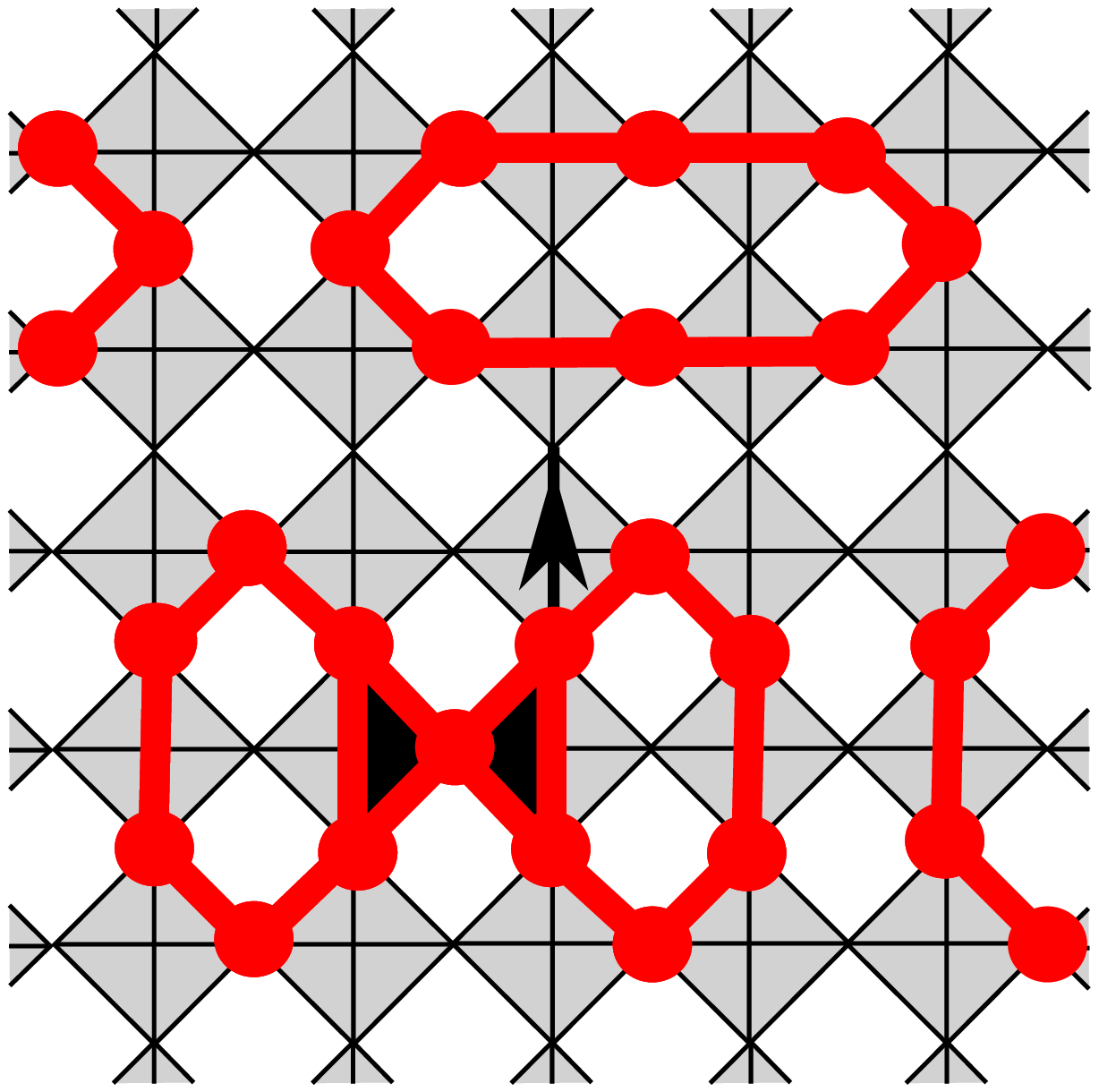}&
(b)~\includegraphics[%
  clip,
  height=22mm,
  keepaspectratio]{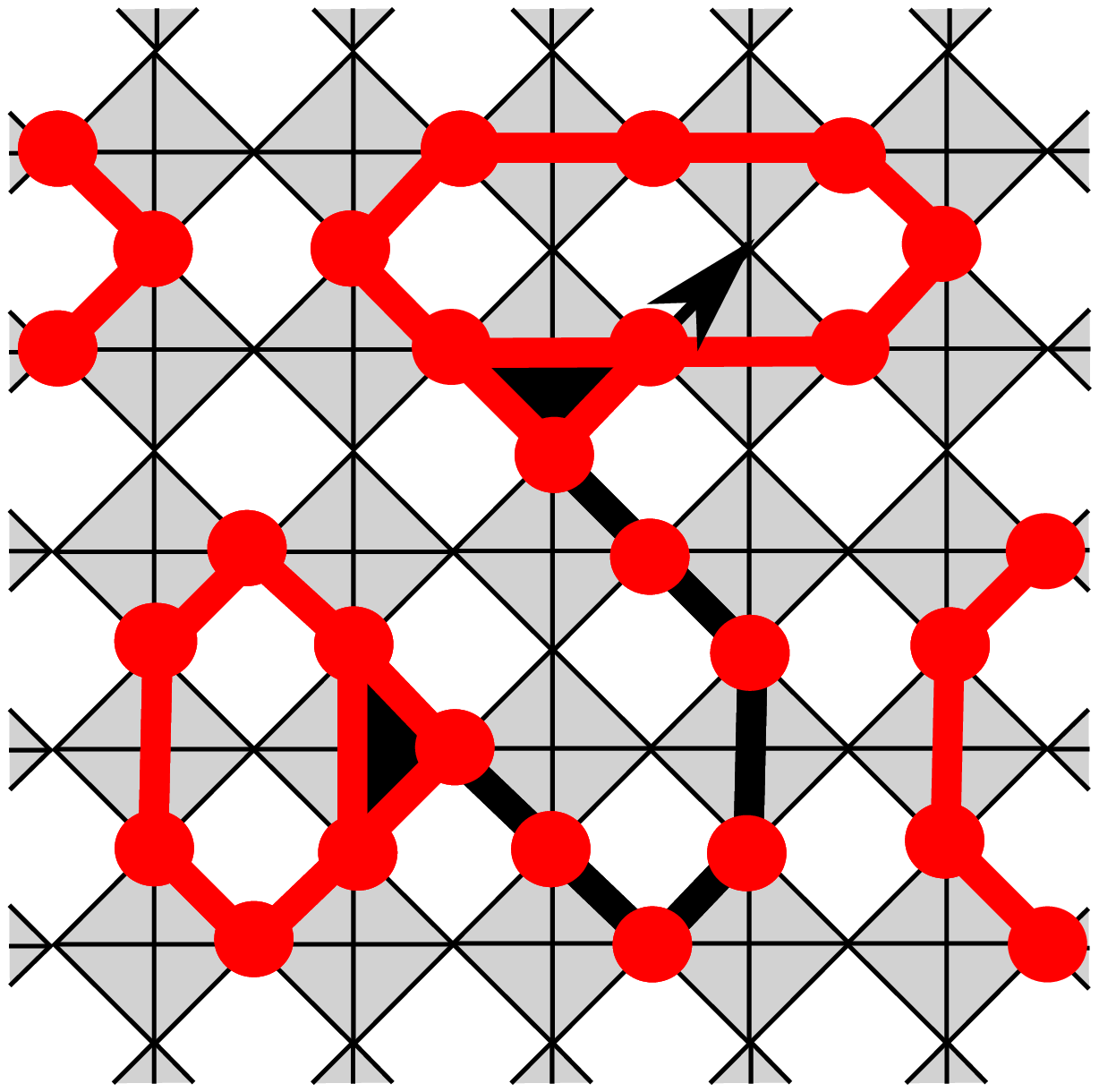}&
(c)~\includegraphics[%
  clip,
  height=22mm,
  keepaspectratio]{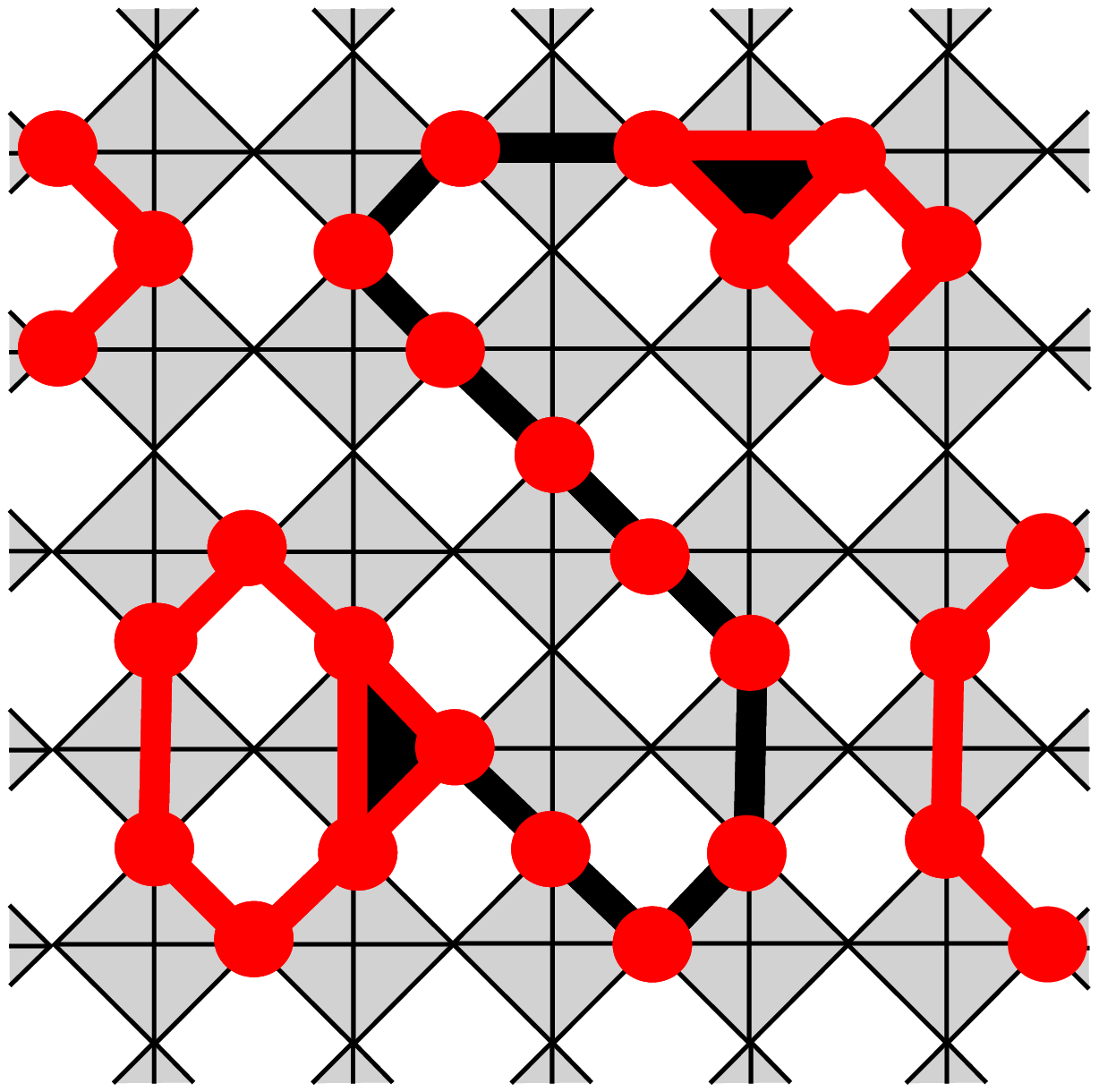}
\end{tabular}\end{center}
\caption{(a) Adding one fermion to the half--filled checkerboard lattice leads
to two defects (marked by black triangles) on adjacent crisscrossed squares. 
(b), (c) Two defects with charge $e/2$  can separate without
creating additional defects. They are connected by a string consisting
of an odd number of fermions.\label{cap:C2_added}}
\end{figure}

Placing one additional particle with charge $e$ onto an empty site
of an allowed configuration leads to a violation of the tetrahedron
rule on two adjacent crisscrossed squares, see Fig.~\ref{cap:C2_added}(a).
The energy is increased by $4V$ since the added particle has four
nearest neighbors (charge gap). There is no way to remove the violations
of the tetrahedron rule by moving electrons. However, fermions on
a crisscrossed square with three particles can hop to another neighboring
crisscrossed square without creating \emph{additional} violations
of the tetrahedron rule, i.e., without increase of the repulsion energy
[see Fig.~\ref{cap:C2_added}(b,c)]. By these hopping
processes, two local defects (violations of the tetrahedron rule)
can separate and the added fermion with charge $e$ breaks into two
pieces. They carry a fractional charge of $e/2$ each. 
In the quantum mechanical case ($t\not=0$), the separation leads
to a lowering of the kinetic energy of order $|t|$. 

Energy and momentum
must be conserved by the decay processes $1e\rightarrow2\mathrm{fcp's}$.
If we associate momentum $\mathbf{k}$ and energy $E(\mathbf{k})$
with the inserted fermion, they must now be shared between the resulting
two fcp's into which it has decayed\begin{equation}
E\left(\mathbf{k}\right)=4V+\epsilon\left(\mathbf{k}_{1}\right)+\epsilon\left(\mathbf{k}_{2}\right).\label{eq:EistEplusE}\end{equation}
Here $\mathbf{k}=\mathbf{k}_{1}+\mathbf{k}_{2}$ and $\epsilon\left(\mathbf{k}\right)$
is the energy dispersion of a fcp. Even though $\epsilon\left(\mathbf{k}\right)$
is at present completely unknown, Eq.~(\ref{eq:EistEplusE}) allows
to predict that for deconfined fcp's the electronic spectral function
should not contain a Fermi--liquid peak, but should show a broad continuum
instead. 

Figure~\ref{cap:C2_added} demonstrates that the two defects can
be thought of as always being connected by a string of occupied sites
consisting of an \emph{odd} number of sites. The fractional charges
can thus be alternatively interpreted as the ends of a string--like
excitation. In this picture, the connection is not static, as two
pairs of fcp's can exchanges partners when the connecting strings
come close.

\paragraph{Quantum Fluctuations.}

\begin{figure}
\begin{center}\begin{tabular}{llll}
(a)~\includegraphics[height=22mm,keepaspectratio]{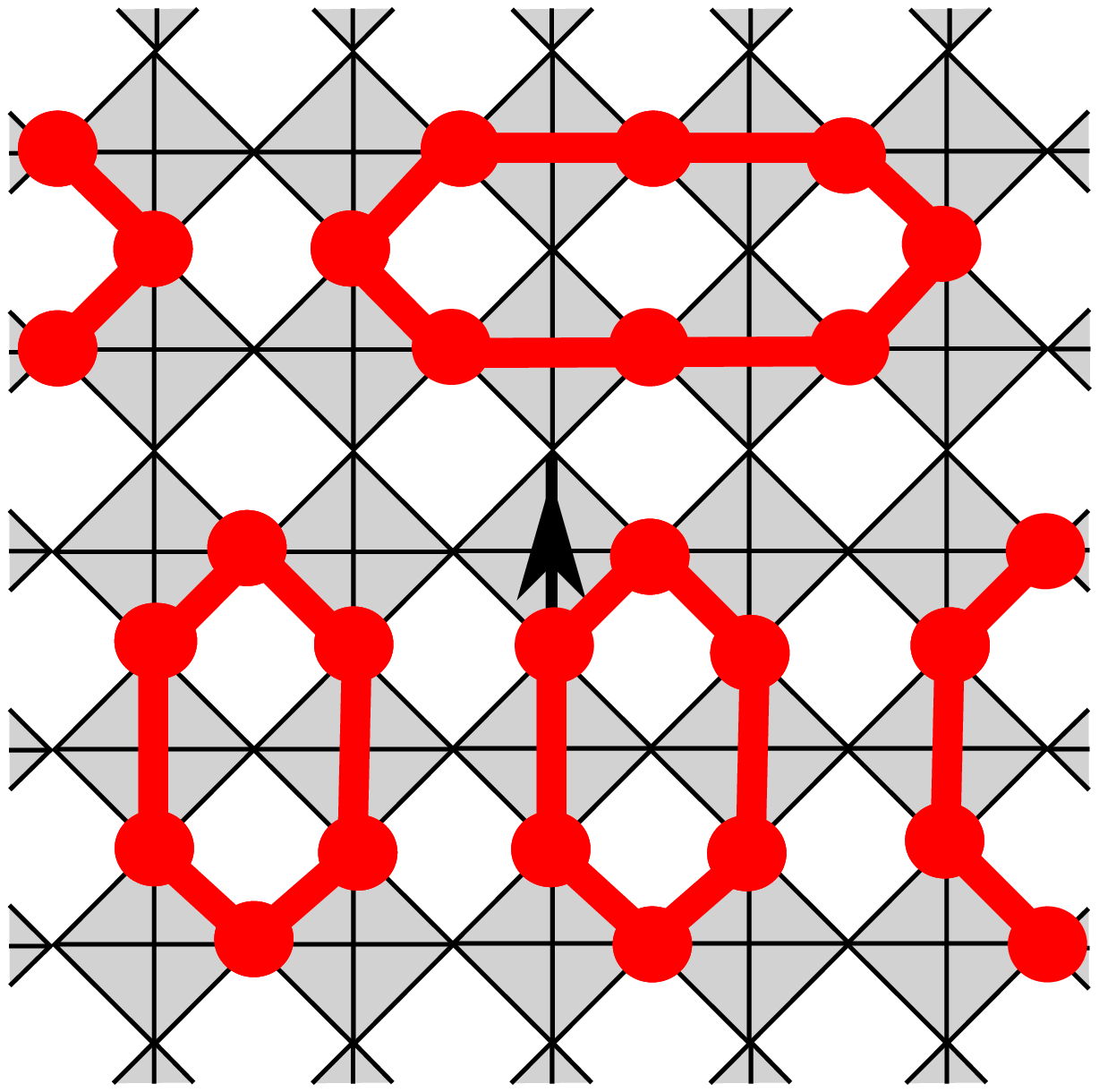}&
(b)~\includegraphics[height=22mm,keepaspectratio]{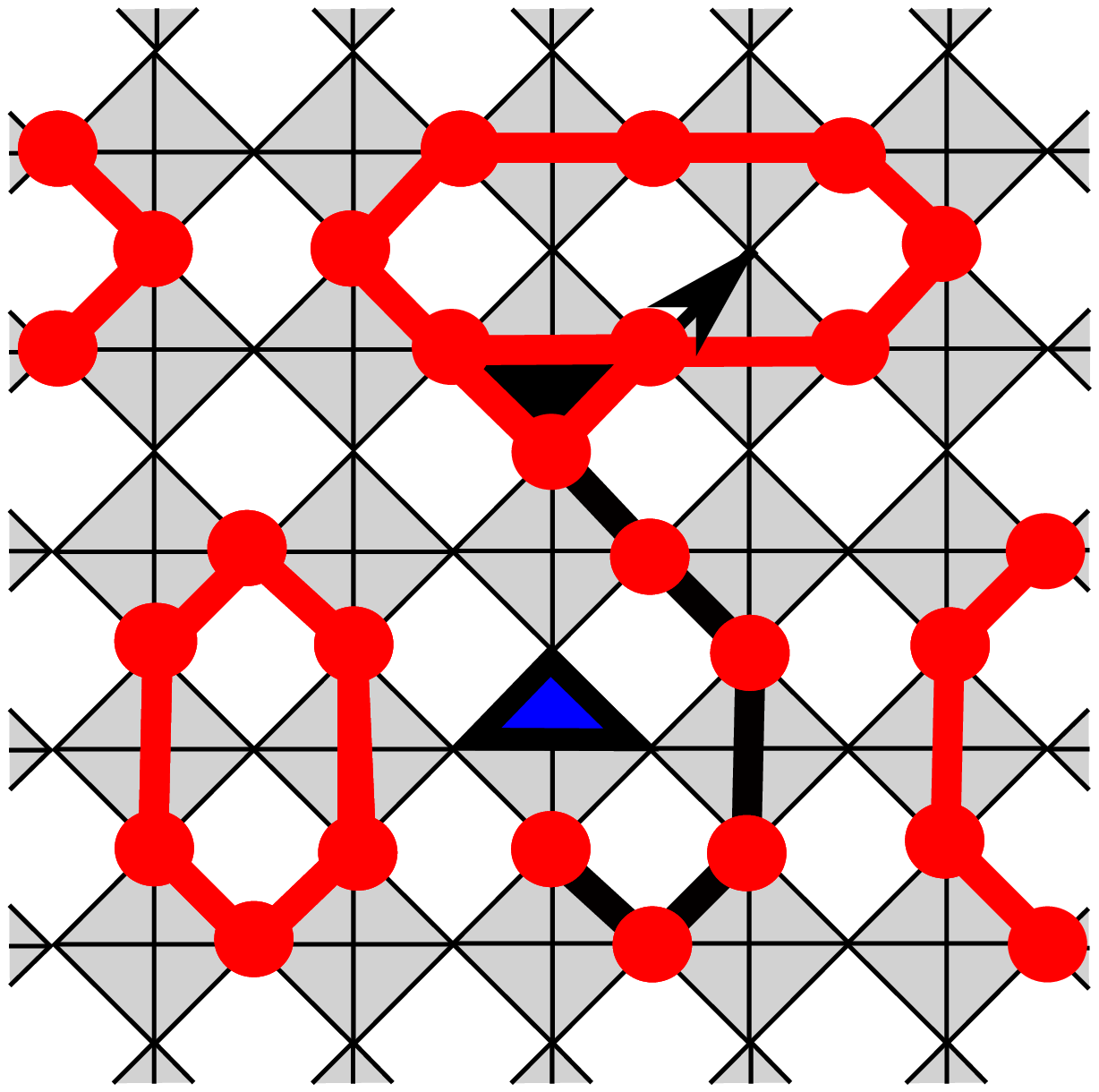}&
(c)~\includegraphics[height=22mm,keepaspectratio]{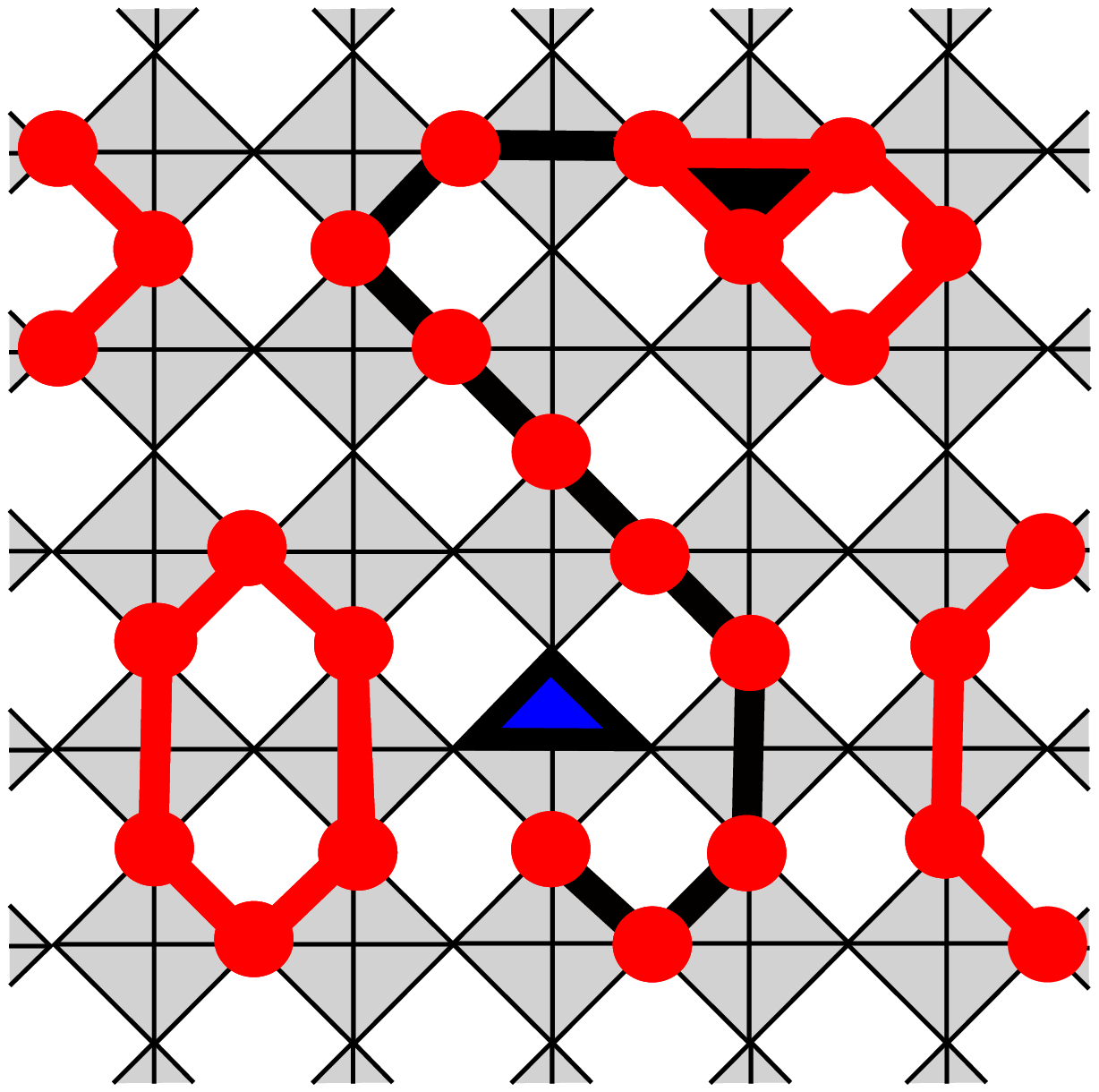}&
(d)~\includegraphics[height=22mm,keepaspectratio]{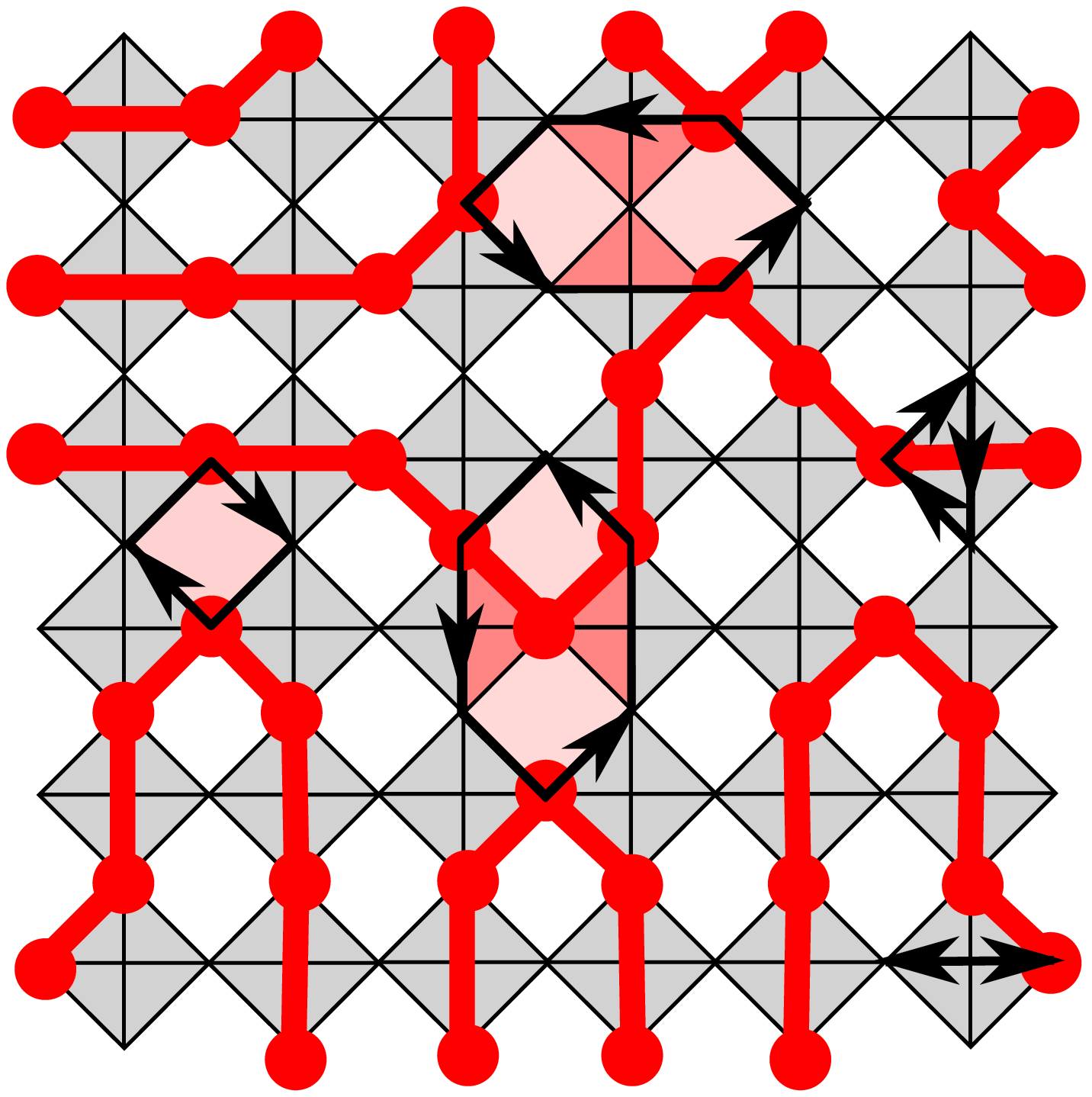}
\end{tabular}\end{center}

\caption{(a)--(c) Hopping of a fermion to a neighboring site in an allowed
configuration generates a fluctuation: (a)~A fractional charged particle
(fcp) and fractional charged hole (fch) are generated. (b)--(c)~The
two defect (marked by triangles) with charge $\pm e/2$ can separate
without creating additional defects and are connected by a string
consisting of an even number of fermions.\label{cap:C2_fluct}
(d) Example of an allowed configuration on a checkerboard lattice
at half filling with possible low--order hopping processes. \label{cap:C4_Checkerboard-lattice}}
\end{figure}
If we relax the constraint of having two fermions on each crisscrossed
square and consider a small but finite ratio $t/V,$ quantum fluctuations
come into play. The quantum fluctuations lead also to fractional charges,
but do not change the net charge of the systems. Starting from an
allowed configuration, the hopping of a fermion to a neighboring site
increases the energy by $V$. One crisscrossed square contains three
fermions while the other has only one fermion, see Fig.~\ref{cap:C2_fluct}(a)--(c).
These so-called vacuum fluctuations (virtual fcp--fch pairs) lead to
two mobile fractional charges with opposite charges $+e/2$ and $-e/2$, which are connected by a string of an \emph{even} number of
fermions. The energy associated with a free fcp and a free fch is
$\Delta E_{\mbox{vac}}=V+\epsilon\left(\mathbf{k}\right)+\bar{\epsilon}\left(-\mathbf{k}\right)$,
where $\bar{\epsilon}\left(-\mathbf{k}\right)$ denotes the kinetic
energy of a fch. Such virtual processes connect different allowed
configurations, lower the total energy, and reduce the macroscopic
degeneracy, as we will see more explicitely in the next section.

\paragraph{Fractional charges in 3D.}

\begin{figure}
\begin{center}\begin{tabular}{cc}
(a)\includegraphics[%
  height=40mm]{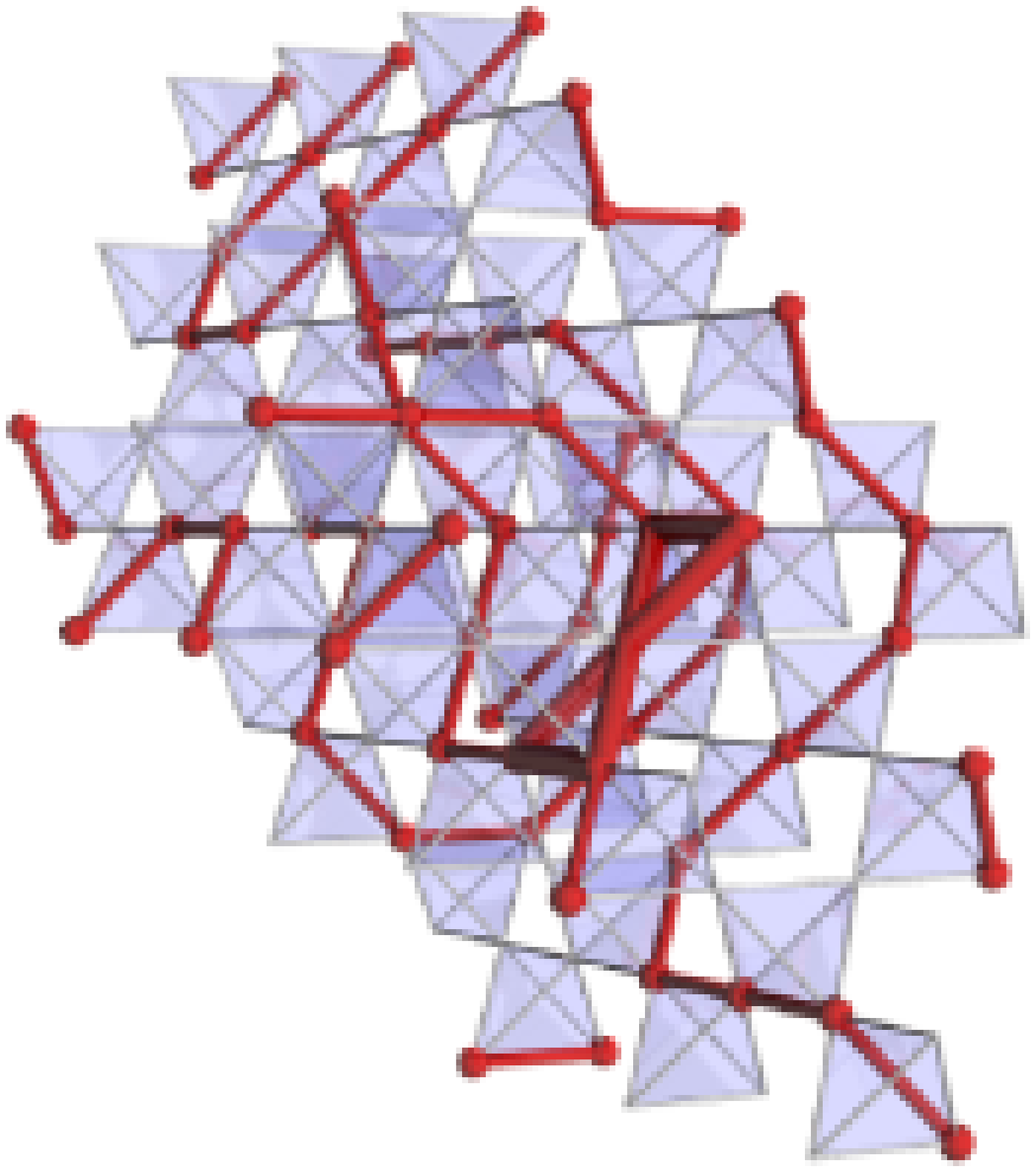}&
~~~
(b)\includegraphics[%
  clip,
  height=40mm]{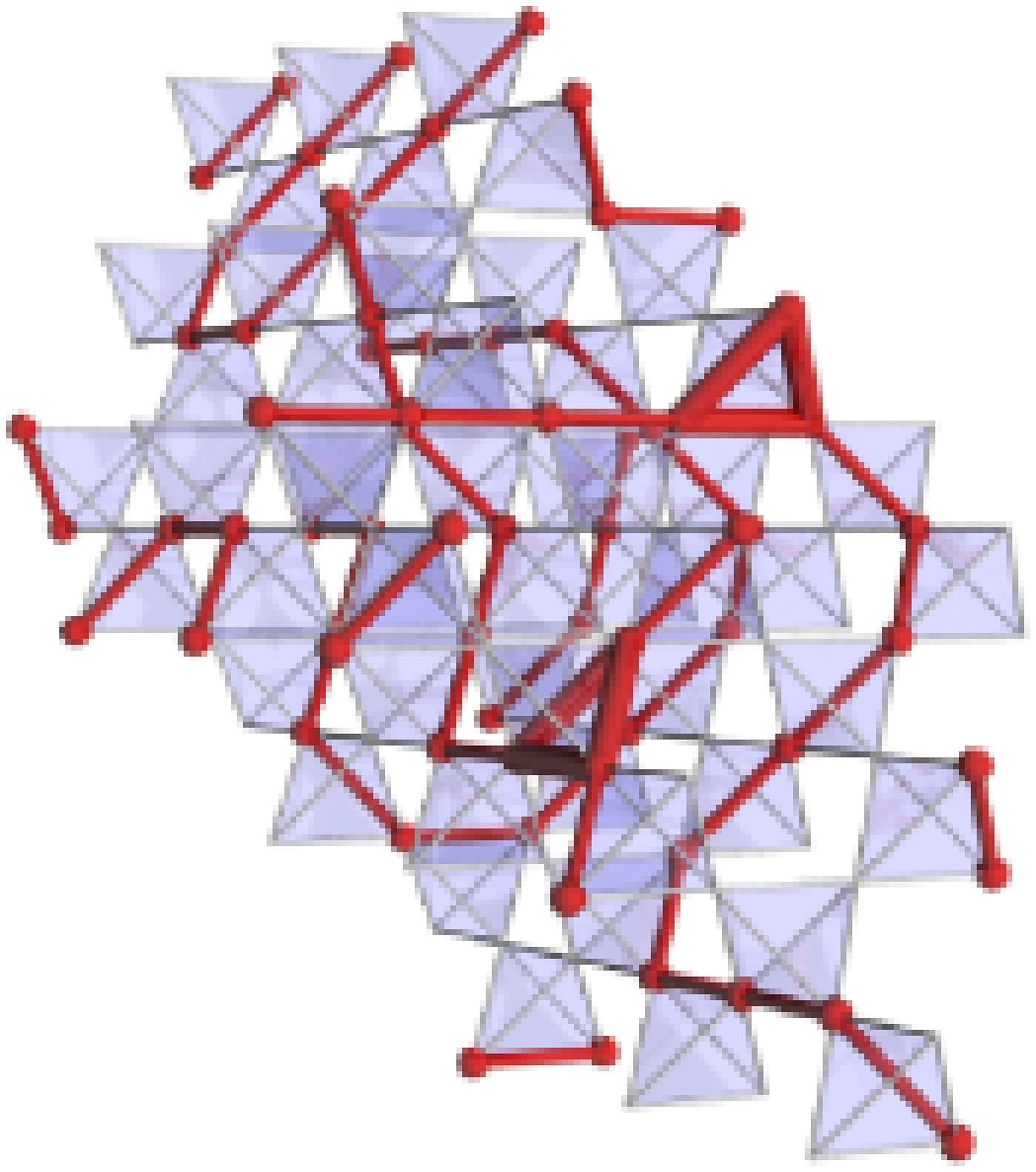}
\end{tabular}\end{center}

\caption{Adding one fermion to the half--filled three-dimensional pyrochlore
lattice leads to two defects on adjacent tetrahedra. The two defects
with charge $e/2$ can separate without creating additional defects
and are connected by a string consisting of an odd number of fermions.\label{cap:C2_pyro}}
\end{figure}
All arguments mentioned above for the existence of fcp's on a 2D checkerboard
lattice can be directly transferred to a 3D pyrochlore lattice at
half filling, see Fig.~\ref{cap:C2_pyro}. Fcp's correspond to tetrahedra
with three fermions and fch's to tetrahedra with only one fermion.

\section{Effective Hamiltonians\label{sec:Effective--Hamiltonian}}

\paragraph{Ring exchange processes in the undoped case.}

The numerical and analytical work is greatly simplified, if a simpler
Hamiltonian $\mathcal{H}$ can be derived that shows in the limit
$|t|\ll V$ the same low--energy physics as the model Hamiltonian (\ref{eq:C2_spinless_hamil}).
A down--folding procedure can be used to define $\mathcal{H}$: Per
definition, it acts only on the subspace of allowed configurations
and includes all virtual processes up to the lowest non--trivial order,
which is $t^{3}/V^{2}$. The relevant processes are shown in see Fig.~\ref{cap:C4_Checkerboard-lattice}(d)
and can be classified as self--energy contributions $H_{\Sigma}$
and ring--exchange processes $H_{\mathrm{eff}}$,
 $\mathcal{H} = H_{\Sigma} + H_{\mathrm{eff}}$. 
The former comprises
the terms which are diagonal in the real space basis. The latter includes
those which connect different allowed configurations. $H_{\Sigma}$
contains fermion hops to an empty neighboring site and back again
as well as hops around an adjacent triangle. These contribute only
a constant, configuration--independent energy shift. It does not
lift the macroscopic degeneracy and will hence be ignored furtheron.
The total amplitude of ring--exchange processes around empty squares
is proportional to $t^{2}/V$. It vanishes for spinless fermions because
the amplitudes for clockwise and counter--clockwise ring--exchange
cancel each other due to fermionic anti--commutation relations. Thus,
the macroscopic ground--state degeneracy is first lifted by ring exchanges
$\sim t^{3}/V^{2}$ around hexagons, and the looked--for effective
Hamiltonian reads\begin{eqnarray}
H_{\mathrm{eff}} & = & -g{\textstyle \sum_{\{\smallhexh,\smallhexv\}}}\big(\,\big|\hexaf\big\rangle\big\langle\hexbf\big|-\big|\hexa\big\rangle\big\langle\hexb\big|+\mbox{H.c.}\,\big),\label{eq:C4_H_eff}\end{eqnarray}
with the effective hopping--matrix element $g=12\  t^{3}/V^{2}>0$ and the
sum taken over all vertical and horizontal oriented hexagons. The
pictographic operators represent the hopping around hexagons which
have either an empty or an occupied central site. The signs of the matrix
elements depend on the representation and the sequence in which the
fermions are ordered. When the sites are enumerated along diagonal rows, an
exchange process commutes an odd number of fermionic operators if
the site in the center of the hexagon is empty and an even number
if the center is occupied. In Ref.~\cite{runge2004}, it has been shown that the effective
Hamiltonian (\ref{eq:C4_H_eff}) gives a good approximation of the
low--energy excitations of the full Hamiltonian (\ref{eq:C2_spinless_hamil})
in the limit considered.

\paragraph{Propagation of defects and extra charges.}
We have derived $H_{\mathrm{eff}}$ as effective Hamiltonian for 
local rearrangement processes as resulting from the (virtual/intermediate)
generation of a defect pair which subsequently recombines, leaving
behind a different allowed generation. For many questions of physical
interest, it is necessary to consider the propagation of defects over
large distances. Analogously to conventional semiconductors, this
refers to long--lived thermally generated defects (e--h pairs
in the semiconductor analogy) as well as to the consequences of slight
doping, i.e., addition of two fcp's by addition of one extra fermion.
The natural generalization of the effective Hamiltonian (\ref{eq:C4_H_eff})
to these cases includes besides the lowest order ring--exchange
processes $H_{\mathrm{eff}}$ a projected hopping term that moves
the defects \begin{eqnarray}
H_{\mathrm{tg}} & = & H_{\mathrm{eff}}-t\sum_{\langle i,j\rangle}P\left(c_{i}^{\dag}c_{j}^{\vphantom{\dag}}+\mbox{H.c.}\right)P.\label{eq:C4_hamil_tg}\end{eqnarray}
The projector $P$ ensures that $H_{\mathrm{tg}}$ acts only on the
subspace of configurations with the smallest possible number of violations
of the tetrahedron rule which is compatible with the number of particles,
e.g., two in the case of one added fermion. We refer to $H_{\mathrm{tg}}$
as the $t$--$g$~model and consider the parameters $t$ and $g$ as independent
(i.e., not restricted to the regime $g\ll t$ as enforced by $t\ll V$). 
This allows to study the effect of ring exchange onto the dynamics
of fractionally charged excitations. In particular, the question of
confinement can be studied even on rather small clusters by increasing
the ring exchange strength $g$ relative to $t$.

\section{Height representation, conserved quantities, and gauge symmetries\label{sec:Height-representation-and}}

\begin{figure}
\begin{center}\begin{tabular}{lll}
\includegraphics[%
  height=25mm,
  keepaspectratio]{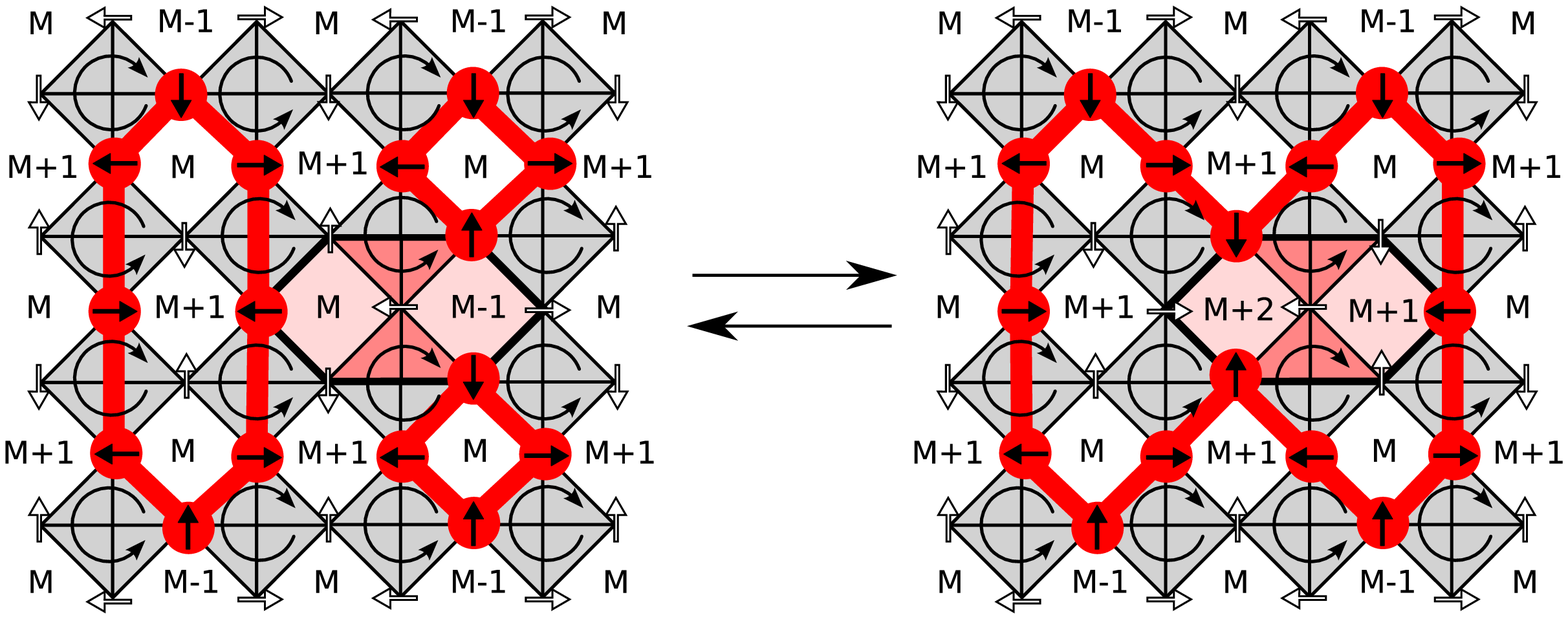}&
\includegraphics[%
  height=25mm]{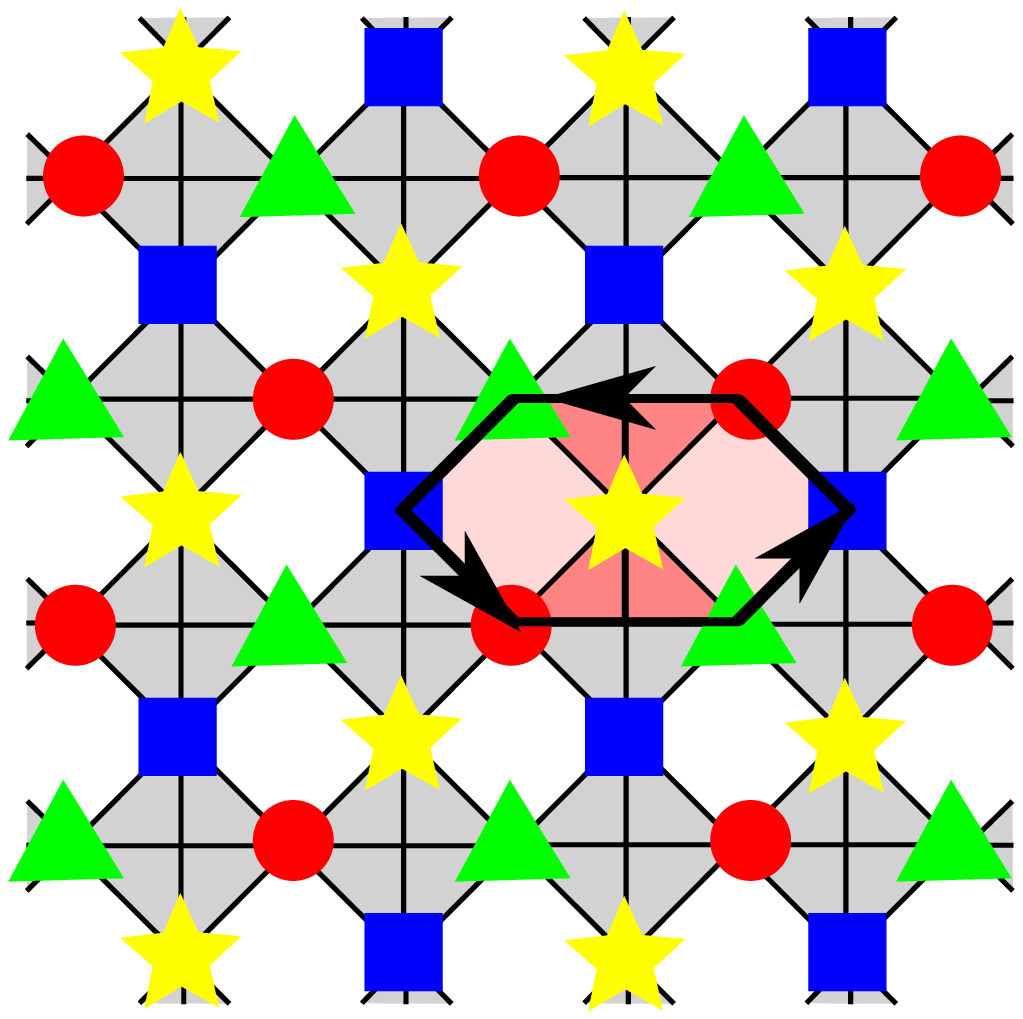}&
\includegraphics[%
  height=25mm]{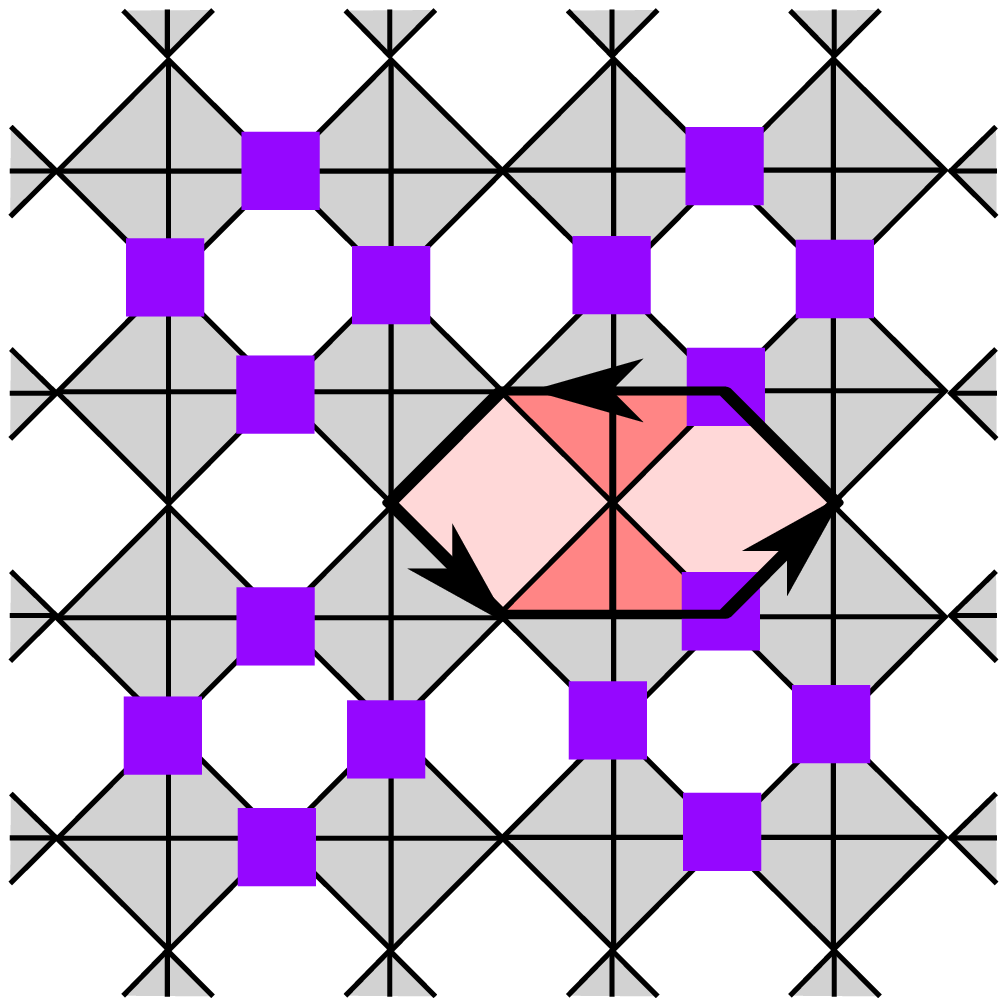}\\
~~(a)&
~~(b)&
~~(c)
\end{tabular}\end{center}

\caption{(a) Height representation for examples of allowed configurations
of a $\sqrt{32}\times\sqrt{32}$ checkerboard lattice with periodic
boundary conditions at half filling. The height field $h$ (numbers
in the non--crossed squares) is uniquely defined for a given configuration
up to an additive constant $M$. The field $\mathbf{f}=\nabla h$
is indicated by small arrows on the lattice sites. Details of the
mapping can be found in the text. The effect on the height fields
of a ring--exchange process around a hexagon is shown explicitely.\label{cap:C4_height}
(b) The effective Hamiltonian conserves the number of fermions on
each of the four sublattices of the checkerboard lattice which are
labeled by {}``blue squares'', {}``red circles'', {}``yellow
stars'' and {}``green triangles.'' (c) Ring--exchange processes
change the number of fermions on sites marked by dark {}``purple squares''
by two.\cite{pollmannDissertation} \label{cap:C4_The-four-sublattices}}
\end{figure}
Conserved quantities of a Hamiltonian allow for a reduction of the
numerical effort by exploiting the resulting block--diagonal form
of the matrix representation. The real space configurations spanning
the subspaces corresponding to these blocks will be referred to as
{}``subensembles.'' Eigenstates can conveniently be classified by
the eigenvalues of the conserved quantities as quantum numbers for
our model. We will now identify several such quantum numbers in order
to characterize different subensembles. A topological quantity, which
is conserved by all local processes, i.e., ring--exchange processes,
is the average tilt of a scalar height field which will be introduced
in the next paragraph. Another useful set of quantum numbers are the
number $(N_{\mathcal{B}},N_{\mathcal{Y}},N_{\mathcal{G}},N_{\mathcal{R}})$
of particles on the four sublattices referred to as blue, yellow,
green and red as shown in Fig.~\ref{cap:C4_The-four-sublattices}(b).
They are conserved by $H_{\mathrm{eff}}$. 

Any allowed configurations of the half--filled checkerboard can uniquely
be represented by a vector field $\mathbf{f}$ for which the discretized
lattice version of the (discrete) curl vanishes, i.e., a pure (discrete)
gradient of a scalar field (height field) $\mathbf{f}=\nabla h$,
see Fig.~\ref{cap:C4_height}. The height field $h,$ is derived
from the local constraint expressed by the tetrahedron rule as follows:
A clockwise or counter--clockwise orientation is assigned alternatingly
to the crisscrossed squares. Arrows of unit length are placed on the
lattice sites. The arrows point along (against) the orientation of
the adjacent crisscrossed squares if the site is occupied (empty).
It is easily checked that allowed configurations are those for which
the discretized line integral of $\mathbf{f}$ around every closed loop
vanishes. Thus, $\mathbf{f}=\nabla h$ defines a height field $h$
up to an arbitrary constant $M$. 

The height at the upper (right) and at the lower (left) boundary of
a finite lattice of $N_{x}\times N_{y}$ squares with periodic boundary
conditions can differ only by an integer $-N_{y(x)}\le\kappa_{y(x)}\le N_{y(x)}$,
which is the same for all columns (rows). This defines topological
quantum numbers $(\kappa_{x},\kappa_{y})$. They remain unchanged
by all \emph{local} processes that transform one allowed configuration
into another, i.e., by ring--exchange processes along contractible
loops. In particular, $H_{\mathrm{eff}}$ merely lowers or raises
the local height of two adjacent plain squares by $\pm2$, as illustrated
in Fig.~\ref{cap:C4_height}(a). We refer to $(\kappa_{x}/N_{x},\kappa_{y}/N_{y})$
as global slope.

The lattice symmetry is broken for states with $(\kappa_{x},\kappa_{y})\ne(0,0)$:\cite{runge2004}
E.g., a finite positive value of $\kappa_{x}/N_{x}$ implies a charge
modulation along diagonal stripes. 
Similarly, a charge density modulation is present 
if the condition $N_{\mathcal{B}}=N_{\mathcal{Y}}=N_{\mathcal{G}}=N_{\mathcal{R}}$
is violated.

For the identification of exactly solvable points in parameter space
and for future quantum Monte Carlo simulations, it would be very advantageous
if all matrix elements had the same sign. As first steps in this direction,
we describe a gauge transformation that changes the global sign of
$g$ in the effective Hamiltonian (\ref{eq:C4_H_eff}) and then show
conditions under which it is possible to remove minus signs for the
half-filled checkerboard lattice.\cite{pollmann2006d} 

Consider a {}``purple'' sublattice $\mathcal{P}$ that contains
the sites around every second plaquette as shown in Fig.~\ref{cap:C4_The-four-sublattices}(c).
Define $\sigma_{{\cal P}}$ as the number of fermions on purple sites
for a given configuration: $\sigma_{{\cal P}}=\sum_{i\in\mathcal{P}}n_{i}$.
One observes that ring--exchange processes change $\sigma_{{\cal P}}$
by two. Thus, if all configurations are multiplied by a factor of $i{}^{\sigma_{\mathcal{P}}}$,
the sign of all ring--exchange matrix--elements are
changed. The invariance with respect to this gauge transformation
proves a global $g\leftrightarrow(-g)$ symmetry. 

\begin{figure}
\begin{center}
\includegraphics[width=80mm]{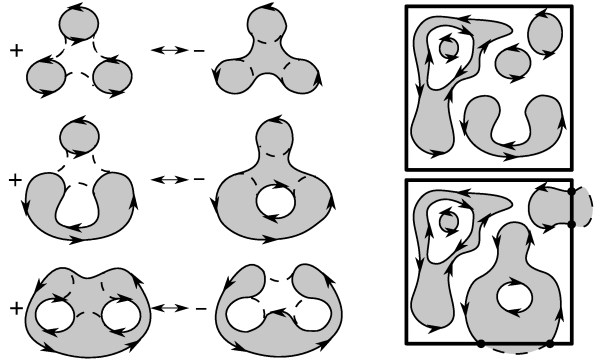}
\end{center}

\caption{The three different actions of the effective Hamiltonian on the topology
of loop configurations are shown in panel (a). Panels (b) and (c)
show representations of two configurations by fully--packed directed
loops.\cite{pollmann2006d} \label{cap:C4_loops}}
\end{figure}
However, a fermionic sign problem remains: Ring--exchange processes
around empty and occupied hexagons carry opposite signs in Eq.~(\ref{eq:C4_H_eff}).
We argue that it can be avoided in certain (but not all) cases. \cite{pollmann2006d}
In order to do so, we represent the ground--state manifold by ensembles
of fully--packed loops as exemplified in Fig.~\ref{cap:C2_allowed}.
We notice that a ring exchange around a hexagon with an occupied center
site does not change the loop topology, whereas a ring exchange around
an empty hexagon always does cause changes in one of the three topological different
ways shown in Fig.~\ref{cap:C4_loops}(a). 

Let us consider allowed configurations with ``fixed'' boundary conditions
with an even number of fermions on the four boundaries. They are represented
by closed loops in the interior and loops terminating at a boundary.
We orient the closed loops as follows: (i) Color the areas separated
by the loops alternatively white and grey, with white being the outmost
color; (ii) orient all loops so that the white regions are always
to the right, see Fig.~\ref{cap:C4_loops}(b). If open loops are
present {[}Fig.~\ref{cap:C4_loops}(c){]}, these are closed arbitrarily
but intersection-free outside the sample and colored as described above. 
Let us assume without loss of generality that the color at infinity is white. We now
notice by inspection of Fig.~(\ref{cap:C4_loops}) that the relative
signs resulting from the exchange processes around empty hexagons
are consistent with multiplying each loop configuration by $i^{r}(-i)^{l},$
where $r$ and $l$ are the total number of the clockwise and counter--clockwise
winding loops, respectively. Hence, by simultaneously changing the sign of the exchange--processes
around empty hexagons and transforming the loop states 
\begin{eqnarray}
|\mathcal{L}\rangle & \rightarrow & i^{l(\mathcal{L})}(-i)^{r(\mathcal{L})}|\mathcal{L}\rangle,\label{eq:C4_gauge_sign_trafo}\end{eqnarray}
we cure the sign problem, thus making the system effectively bosonic. 

This construction need not work for periodic boundary
conditions: Firstly, only even--winding subensembles (sectors) on
a torus allow for such a two--color coloring. Secondly, even then it might
be possible to dynamically reverse the coloring while returning to
the same loop configuration. However, the exact diagonalization results
presented below (see Fig.~\ref{cap:C4_Ground-state-energy_HF}) suggest
that for periodic boundary conditions on \emph{even} tori (preserving
the bipartiteness of the lattice), the lowest--energy states belong
to a sector where such a transformation works. We remark that the
presented non--local loop--orienting construction is restricted to
the effective Hamiltonian (\ref{eq:C4_H_eff}), i.e., to the ring--exchange
processes of length six.

\section{Ground states and lowest excitations in the undoped case}

In order to discuss the possible confinement of fcp's, we have to
investigate the nature of the quantum--mechanical ground state of
the undoped system. We do so in the approximation of an effective
ring-exchange Hamiltonian (\ref{eq:C4_H_eff}) acting only on the
subset of allowed configurations. Following Rokhsar and Kivelson,
\cite{rokhsar1988} we add to $H_{\mathrm{eff}}$ an extra term that
counts the number of flippable hexagons. The extended Hamiltonian
reads 

\begin{eqnarray}
H_{\mathrm{g}\mu} & = & H_{\mathrm{eff}}+\mu{\textstyle \sum_{\{\smallhexh,\smallhexv\}}}\big(|\hexafg\rangle\langle\hexafg|+|\hexbfg\rangle\langle\hexbfg|\big),\label{eq:C4_H_gmu}\end{eqnarray}
 where the pictographic operators with grey--colored dots in the center
are summed over all flippable hexagons, independent of the occupancy
of the site in the center. Next, we discuss some limiting cases of
the Hamiltonian (\ref{eq:C4_H_gmu}):\\
\begin{figure}
\begin{center}\begin{tabular}{ccc}
(a)~\includegraphics[%
  height=25mm,
  keepaspectratio]{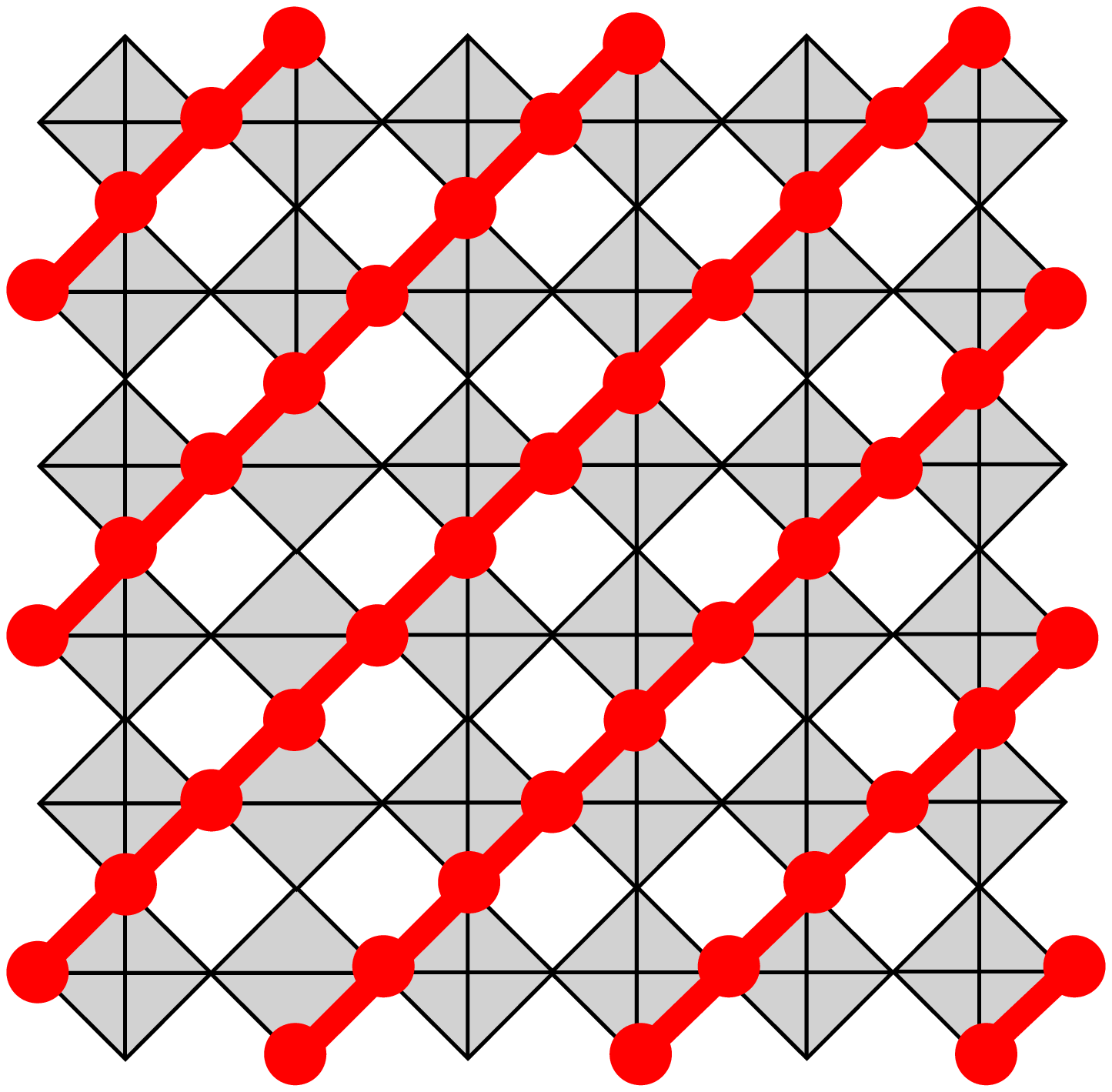}&
(b)~\includegraphics[%
  height=25mm,
  keepaspectratio]{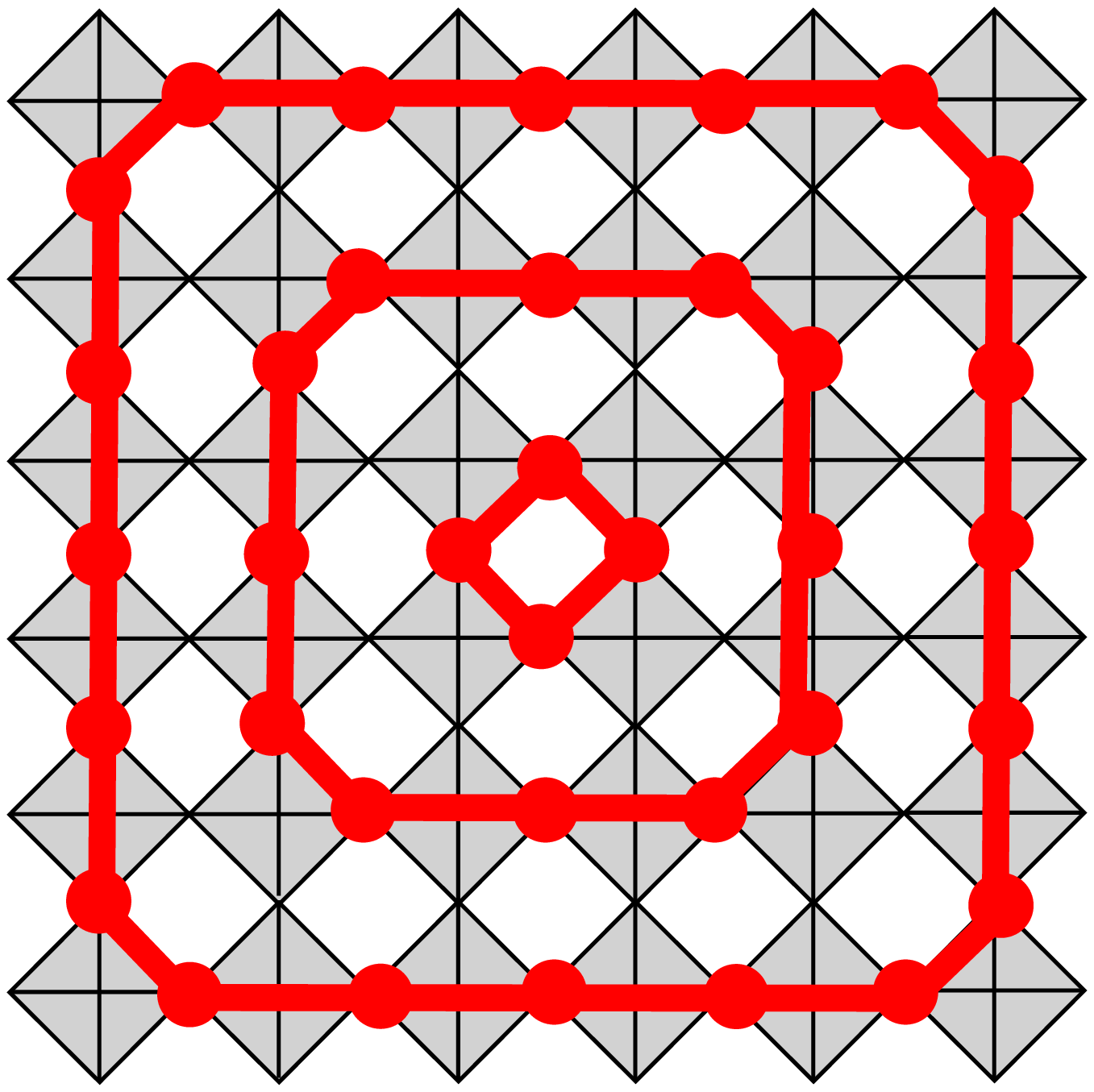}&
(c)~\includegraphics[%
  height=25mm]{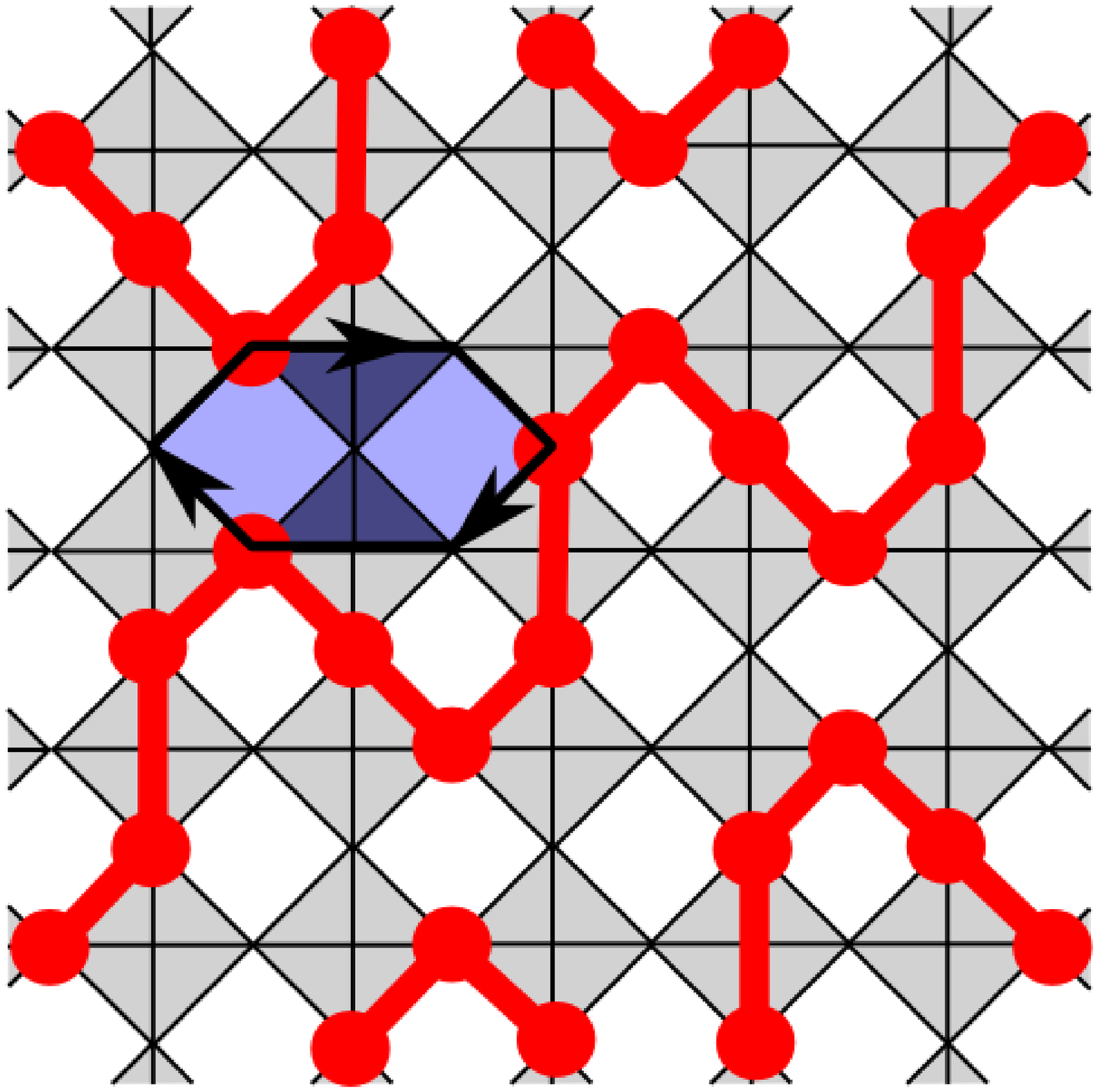}
\end{tabular}\end{center}

\caption{(a)--(b) Fragments of possible ``frozen'' ground states (no flippable
hexagons). \label{cap:C4_frozen} (c) One of several configurations of the half--filled checkerboard
lattice that maximize the number of flippable hexagons. The unit cell
contains $\sqrt{20}\times\sqrt{20}$ sites. \label{cap:C4_flip_f2}}
\end{figure}
(i) $\mu\rightarrow+\infty$: All configurations which contain no
flippable hexagons (frozen configurations) become ground states with
$E=0$. Some are shown as Fig.~\ref{cap:C4_frozen} (a,b). \\
(ii) $\mu\rightarrow-\infty$: Ground states are configurations with
maximal number of flippable hexagons $N_{\mathrm{fl}}$. Using
a simple Metropolis--like Monte Carlo algorithm for lattices with up to 1000 sites, 
we always find that configurations of the type
shown in Fig.~\ref{cap:C4_flip_f2}(c) or slight variations thereof
maximize $N_{\mathrm{fl}}$. Such configurations will be referred
to as {}``squiggle'' configurations.\cite{penc2006} Our numerical
results suggest that in the thermodynamic limit the degenerate ground--state
configurations all lie in the $(\kappa_{x},\kappa_{y})=(0,0)$ subspace.
\\
(iii) $\mu=g>0$: This is the exactly solvable Rokhsar--Kivelson (RK)
point. \cite{rokhsar1988} Following the original RK construction,
we rewrite the Hamiltonian (\ref{eq:C4_H_gmu}) for $\mu=g$ in a way
which explictly shows that some liquid--like ground states have energy $E=0$ and, thus, become degenerate
with the frozen states, i.e., the $\mu$$\rightarrow$$+$$\infty$ solutions.
We assume that we can use the gauge transformation (\ref{eq:C4_gauge_sign_trafo})
to change the sign in the second term and to rewrite the Hamiltonian
as\begin{eqnarray}
H_{\mathrm{g}=\mu} & = & g\sum_{\{\smallhexh,\smallhexv\}}\big[\big(|\!\hexafg\!\rangle-|\hexbfg\!\rangle\big)\times\big(\langle\!\hexafg|-\langle\hexbfg\!|\big)\big].\label{eq:C4_H_RK_II}\end{eqnarray}
Since this is a sum over projectors, all eigenvalues are non--negative.
Furthermore, after re--gauging, \emph{all} off--diagonal elements are
non--positive and for each subensemble $\ell$ an exact ground--state
wavefunction is given by the equally weighted superposition of its
configurations $|C_i^{(l)}\rangle$, i.e., $|\psi_{0,RK}^{(l)}\rangle\sim\sum_{i}|C_{i}^{(l)}\rangle$.
These coherent superpositions are the analogs of the Resonating Valence
Bond (RVB) state, originally discussed by L. Pauling\cite{pauling1953} and 
P. W. Anderson.\cite{anderson1973}
Their energy is easily computed: Each flippable hexagon
contributes $-g$ and $+\mu$, thus for $g=\mu$: $\langle\psi_{0,RK}^{(l)}|H_{\mathrm{g}\mu}|\psi_{0,RK}^{(l)}\rangle=0.$
Note that for fermionic systems, we find a well defined RK point only
if a gauge transformation exists such that all off--diagonal matrix
elements are non--positive. Otherwise the energy is most likely larger
than zero and the subensembles consequently do not form a ground state. 

\begin{figure}
\begin{center}(a)\includegraphics[%
  height=42mm]{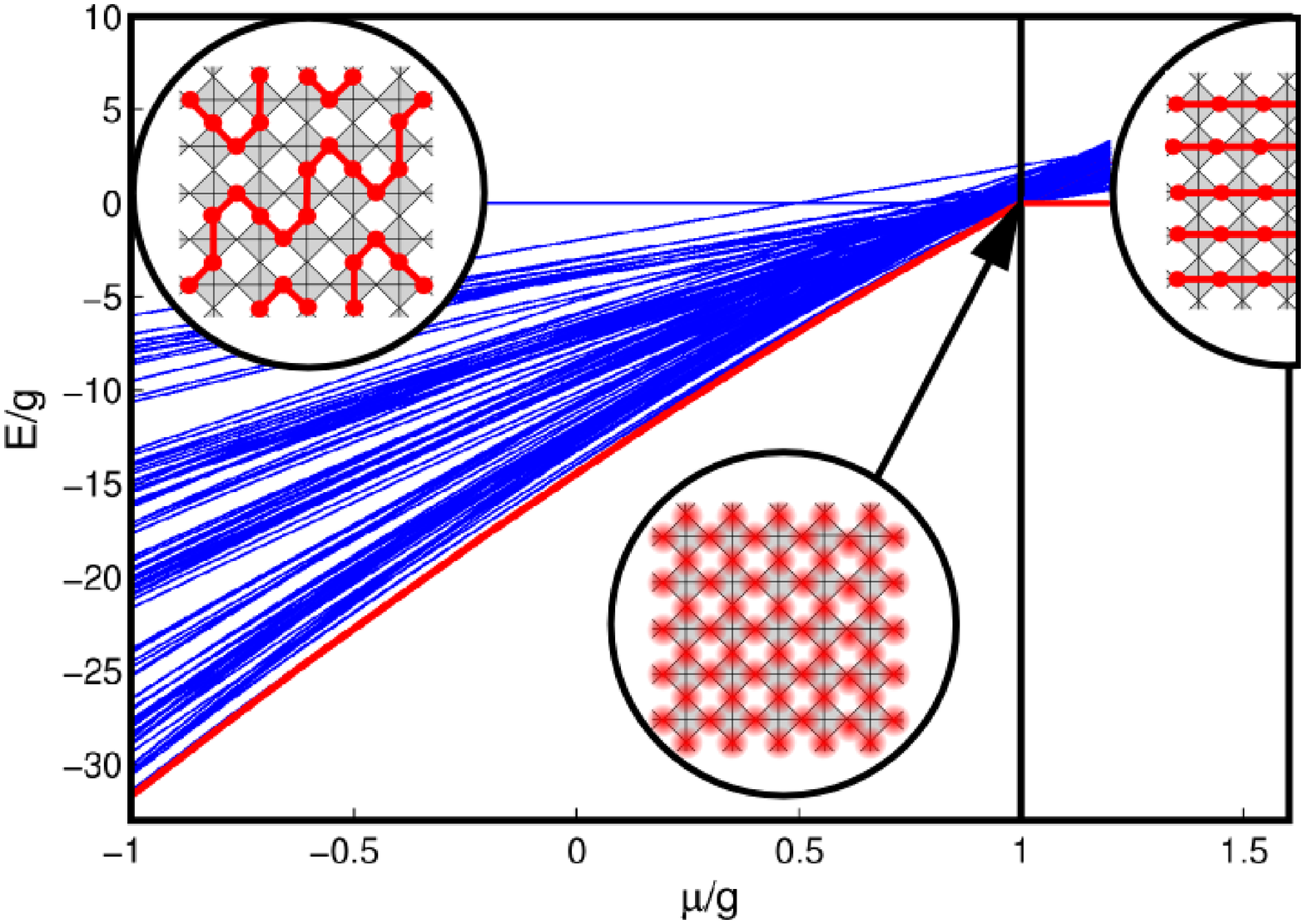}~~(b)\includegraphics[%
  height=42mm]{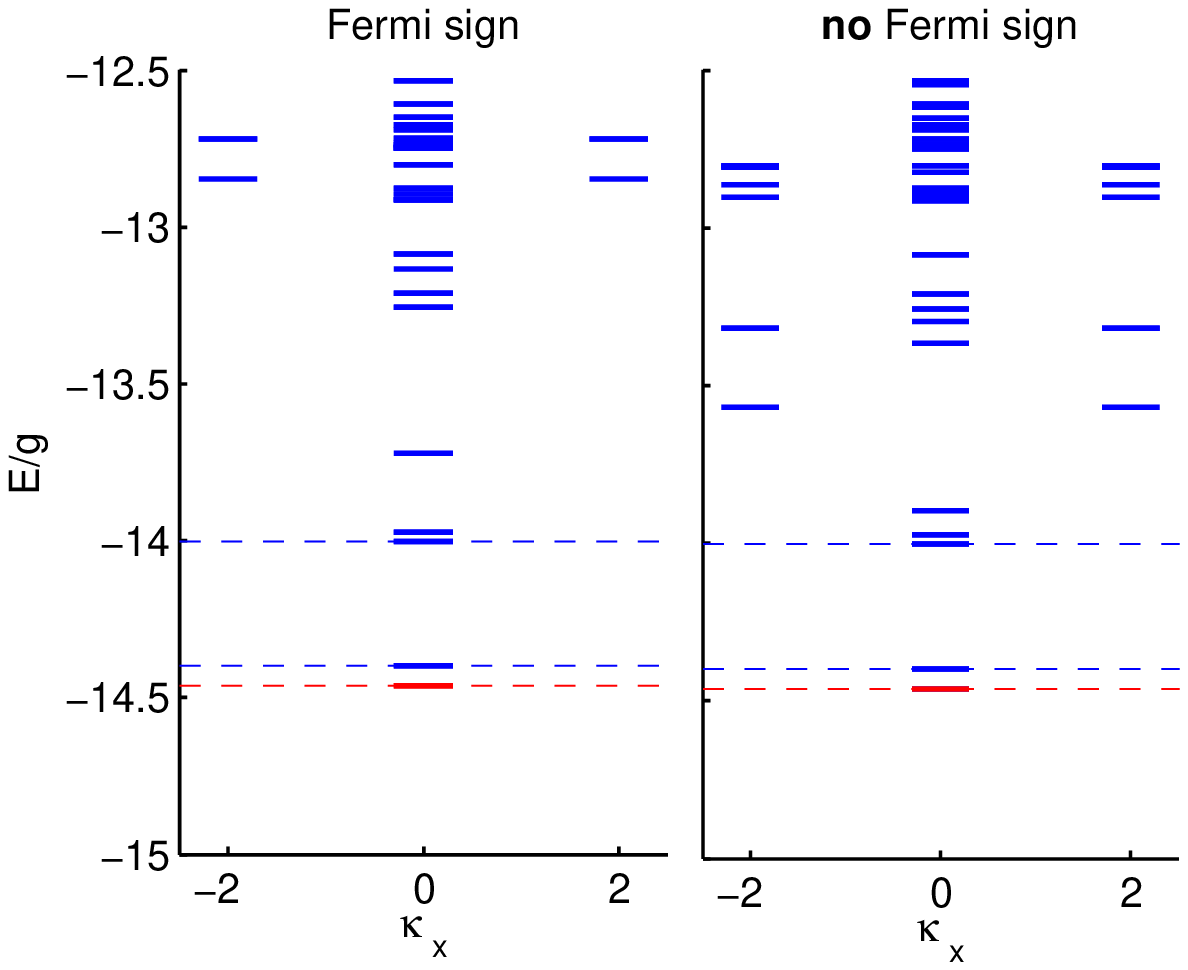}~\end{center}

\caption{(a) Energies of the ground state and lowest excited states in each
subensemble of a 72--site half--filled checkerboard cluster for different
values of $\mu$ of the $g$--$\mu$ Hamiltonian. Level crossing of ground
states occurs only at $\mu=g$. The insets indicate different phases:
Maximal flippable plus fluctuations for $\mu<g$, a critical point
$\mu=g$ where the ground state is an equally weighted superposition
of all configurations, and frozen configurations as ground states
for $\mu>g$.\label{cap:C4_Energies-gmu}
(b)~Left side: Ground--state
energy and energies of the lowest excited states of the effective
Hamiltonian $H_{\mathrm{eff}}$ in subspaces with different global
slopes $\left(\kappa_{x},\kappa_{y}\right)$ $=\left(\kappa_{x},0\right)$
for 72--site cluster. Right side: Same system, but assuming same signs
for all matrix elements ({}``bosonic calculation'').\cite{pollmann2006d}\label{cap:C4_Ground-state-energy_HF}}
\end{figure}

Next, we explore by means of numerically exact diagonalization the eigenstates of
small clusters general $\mu$ values. For the actual calculations,
we first generate all configurations that fulfill the tetrahedron
rule, then group them according to quantum numbers and generate a
sparse block--diagonal matrix representation of the Hamiltonian (\ref{eq:C4_H_gmu}).
For a 72--site checkerboard cluster with periodic boundary conditions,
the $16\ 448\ 400$ dimensional low--energy Hilbert space of allowed
configurations can be decomposed into a few hundred subspaces, where
the largest one has $1\ 211\ 016$ dimensions. The low--energy states
of the sparse block--diagonal Hamiltonian matrix and physical properties
such as charge density distribution and density--density correlations
can easily be obtained on a 64--bit workstation. 
Figure~\ref{cap:C4_Energies-gmu}(a)
shows energies of the ground-state and the lowest excited states
of all subensembles. 

(iv) For $\mu<g$, two ground states are found in subensembles with
$(\kappa_{x},\kappa_{y})=(0,0)$ and $(N_{\mathcal{B}},N_{\mathcal{Y}},N_{\mathcal{G}},N_{\mathcal{R}})=(6,6,12,12)$
and $(12,12,6,6)$, respectively. These are superpositions of configurations
with the maximal number of flippable hexagons. At the physical point
$\mu=0$, the 72--site system is in a crystalline and confining phase.
In the thermodynamic limit, we expect to recover the 10--fold degeneracy
of the squiggle phase instead of the two--fold degeneracy. 
For small values $\mu>0$, the average
weight of configurations in the ground state with the maximum number
$N_{\mathrm{fl}}$ is large (not shown), but decreases with increasing
$\mu$ until at the RK point, i.e., $\mu$=$g$, the ground state is formed
by an equally weighted superposition of all configurations of a certain
subensemble. 

(v) For $\mu>g$, we find essentially the same properties as described
above in the limit $\mu\rightarrow\infty.$

In summary, we find a confining phase and a phase in which the ground states are
given by static isolated configurations. The two phases are separated
by a point with deconfined excitations, i.e., the Rokhsar--Kivelson
(RK) point. \cite{rokhsar1988} The main finding is that the original
effective Hamiltonian is in a confining phase with a long--range ordered
ground state (squiggle phase). This phase maximizes the gain in kinetic
energy and is stabilized by quantum fluctuations (order from disorder).
The results of the exact diagonalization on small samples indicate
that the symmetry remains broken all the way along the $\mu$--axis
up to the RK point. This observation is also strongly disfavoring
a deconfining phase to the left of the RK point. Hence, the fermionic
RK point is likely to be an isolated quantum critical point just as
it is for the bosonic model.\cite{shannon2004} 

We come back to the fermionic sign problem.  Figure~\ref{cap:C4_Ground-state-energy_HF}(b)
compares energies of a system in which the Fermi sign is taken into
account to those from calculations which exclude the Fermi sign. The
ground--state energy, the first excited states in the $(\kappa_{x},\kappa_{y})=(0,0)$
sector, and the weights of the different configurations in the corresponding
eigenstates are the same in both case. In some subensembles
with $(\kappa_{x},\kappa_{y})\ne(0,0)$ the energies of the ground
states including the Fermi sign are higher than the ones for bosons
(no Fermi sign). 

\begin{figure}
\begin{center}\begin{tabular}{lll}
(a)~\includegraphics[%
  height=30mm]{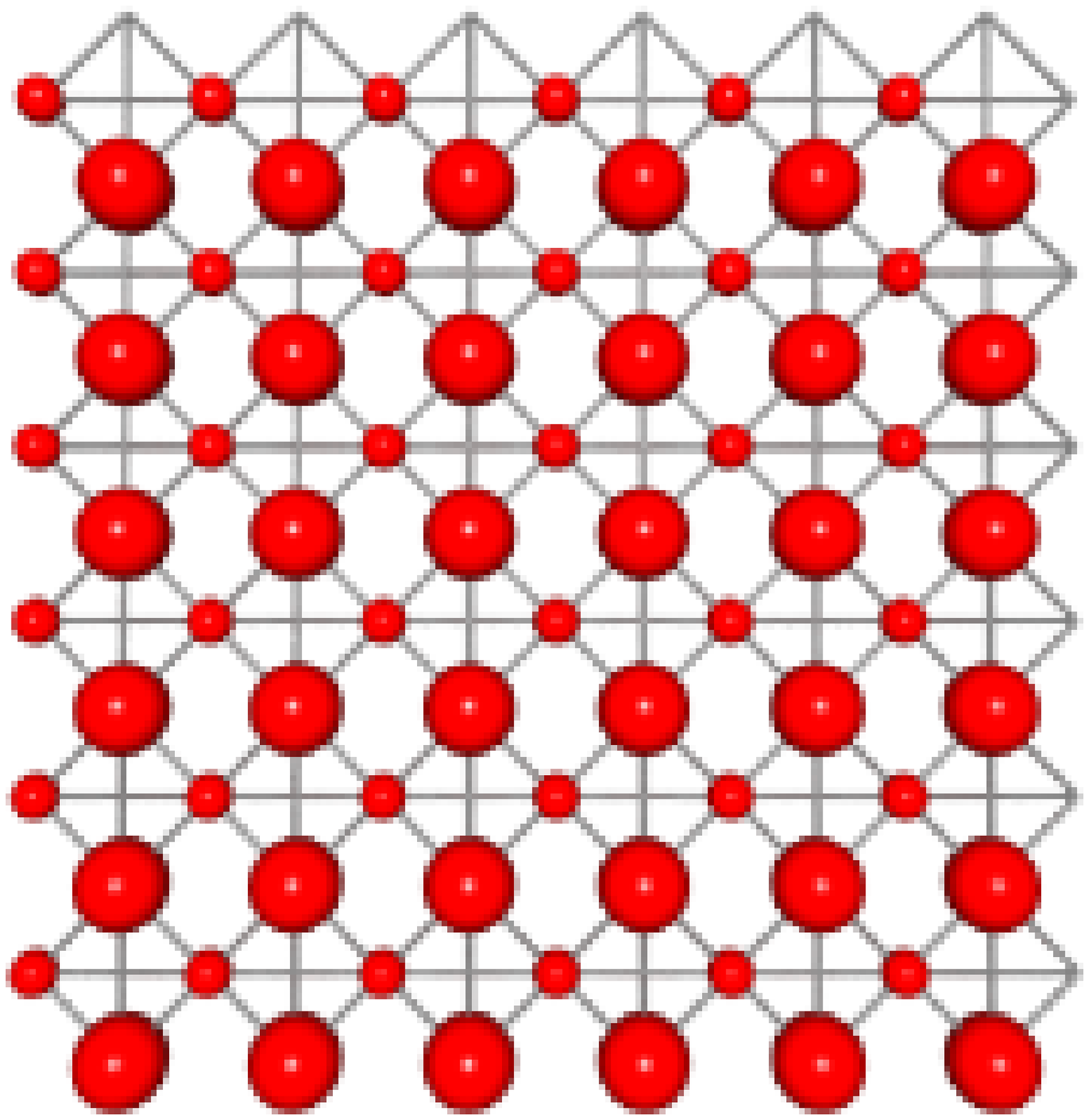}&
(b)~\includegraphics[%
  height=30mm]{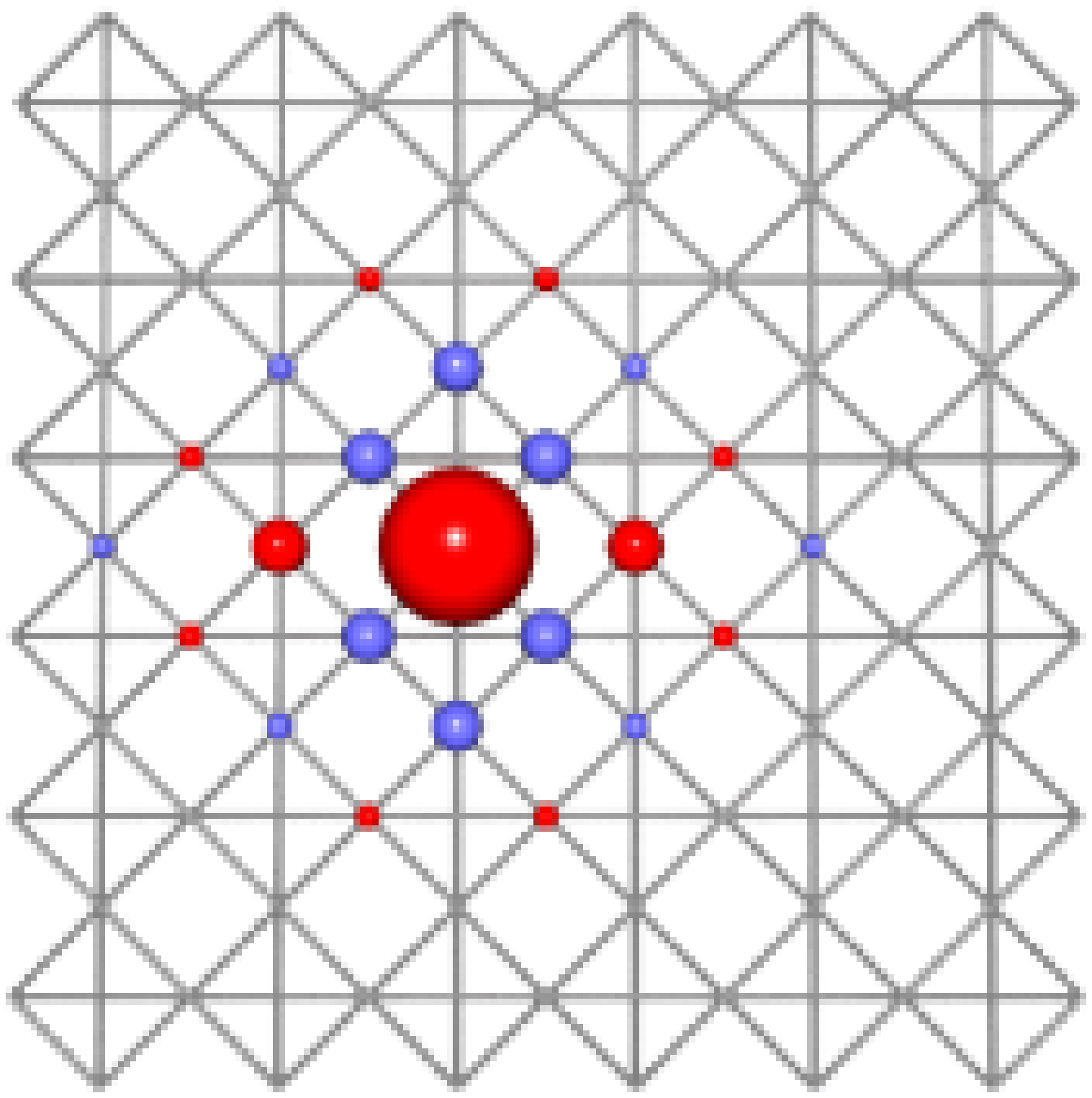}&
(c)~\includegraphics[%
  height=30mm]{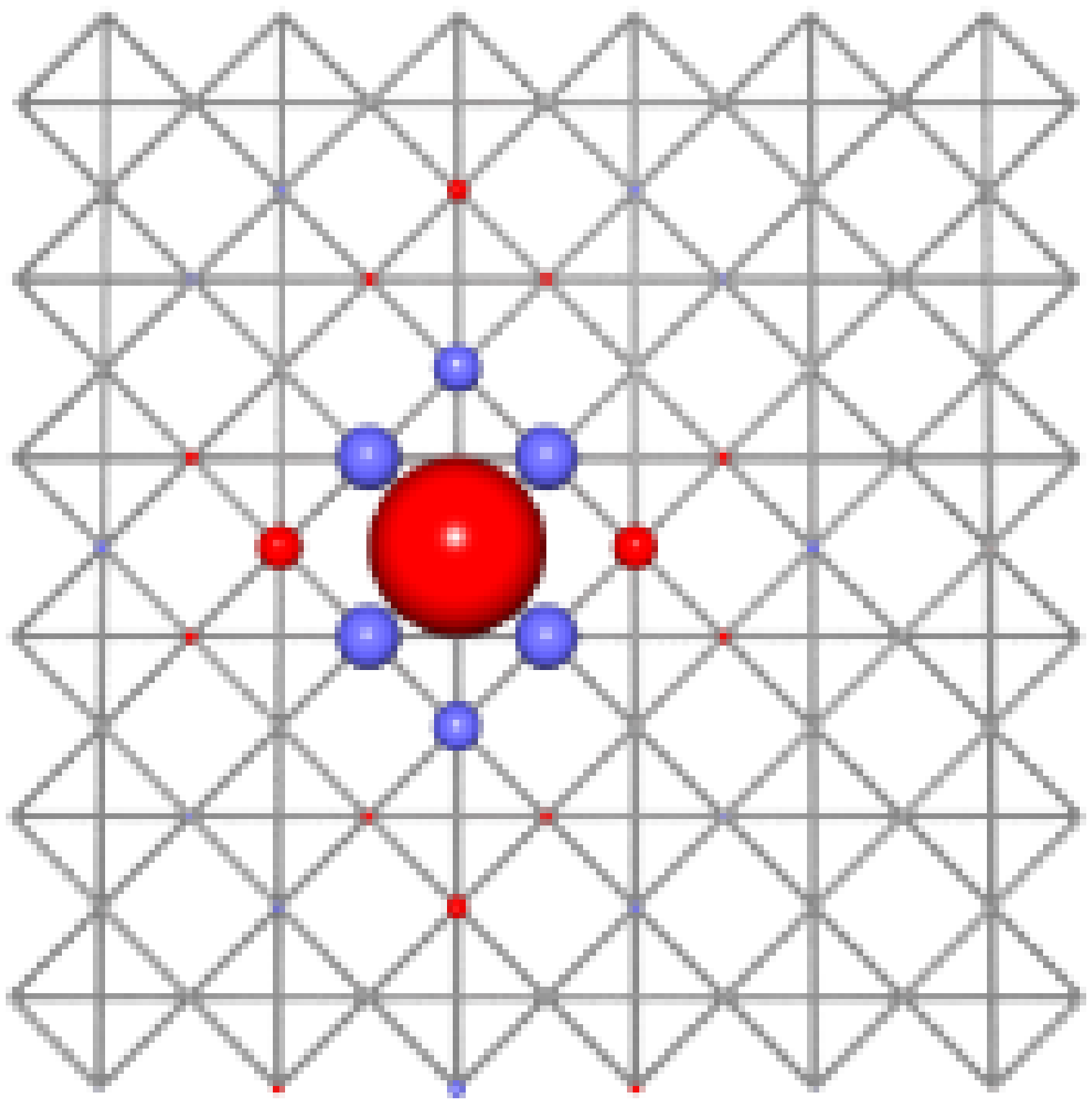}
\end{tabular}\end{center}

\caption{Half filling: (a) Charge density distribution for one of the two
ground sates. (b) Corresponding density--density correlation function
$C_{i_{0}i}^{(l)}$. 
The site $i_{0}$ with average density 2/3 shows
up as the largest dot in the panels (b)--(c). The radius of the dots
is proportional to the absolute value. A red or blue color represents
a positive or negative value, respectively. (c) Classical density--density
correlation function. \label{cap:C4_Panels_nc_HF} }
\end{figure}

At the physical point $\mu=0,$ the 72--site ground state is two-fold
degenerate with quantum numbers $(N_{\mathcal{B}},N_{\mathcal{Y}},N_{\mathcal{G}},N_{\mathcal{R}})=(6,6,12,12)$
or $(12,12,6,6)$. 
The resulting charge order with alternating stripes of average occupation
 $1/3$ and $2/3$ is shown in Fig.~\ref{cap:C4_Panels_nc_HF}(a).
The density--density correlation function in the quantum--mechanical
ground states $|\psi_0^{(l)}\rangle$ {[}Fig.~\ref{cap:C4_Panels_nc_HF}(b){]} \begin{equation}
C_{i_{0}i}^{(l)}=\langle\psi_{0}^{(l)}|n_{i}n_{i_{0}}|\psi_{0}^{(l)}\rangle-\langle\psi_{0}^{(l)}|n_{i}|\psi_{0}^{(l)}\rangle\langle\psi_{0}^{(l)}|n_{i0}|\psi_{0}^{(l)}\rangle.\label{eq:C4_density--density}\end{equation}
is best understood as reflecting the algebraic correlations present
in the average over all classically degenerate configurations, see
Fig.~\ref{cap:C4_Panels_nc_HF}(c) and Ref.~\cite{pollmann2006a}.
Specific quantum--mechanical features resulting from ring hopping
become visible in the difference of the actual correlations and have
been discussed in Ref.~\cite{runge2004}. 

\begin{figure}
\begin{center}
\begin{tabular}{ccc}
(a)\includegraphics[%
  width=27mm]{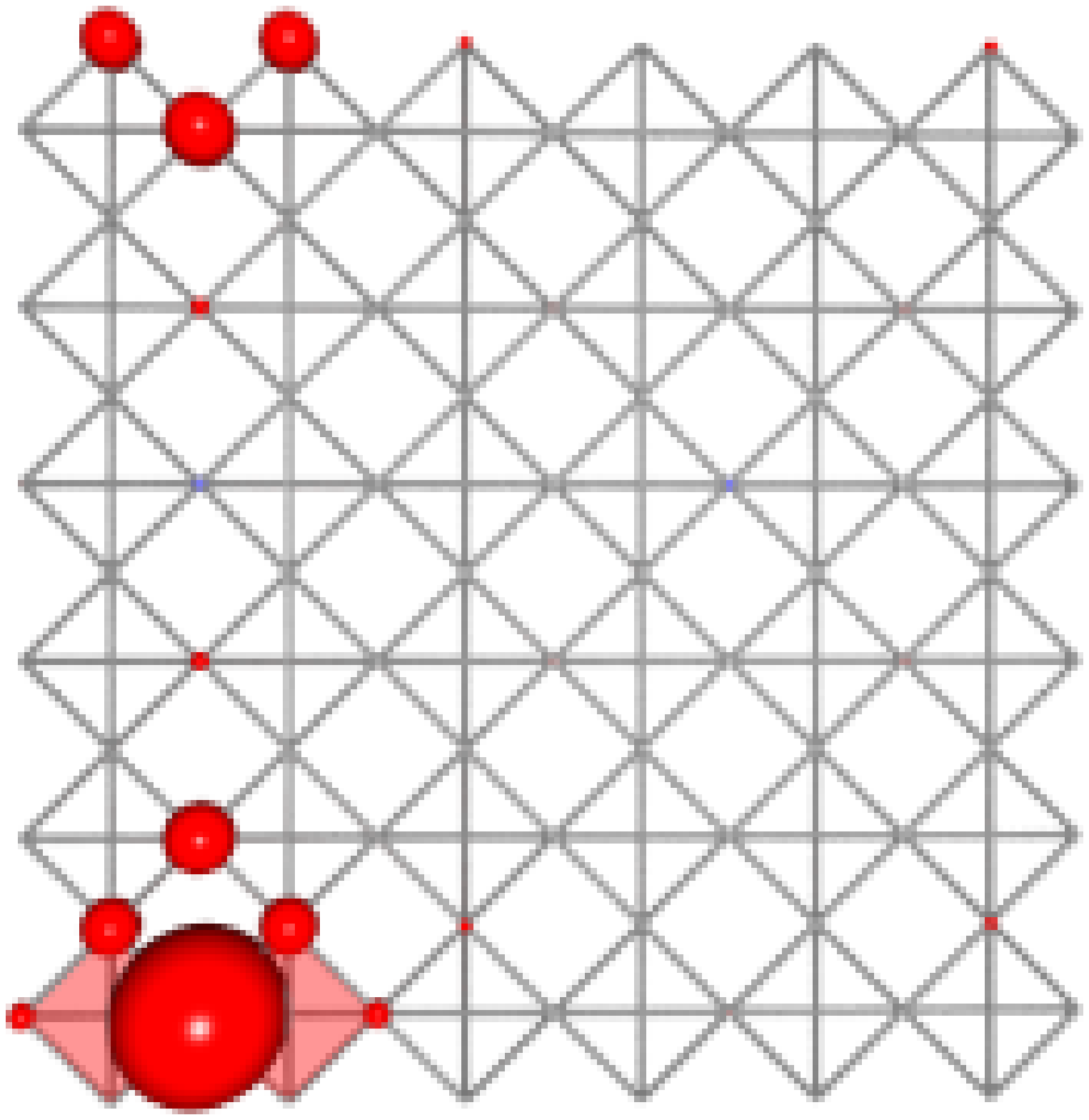}&
(b)\includegraphics[%
  width=27mm]{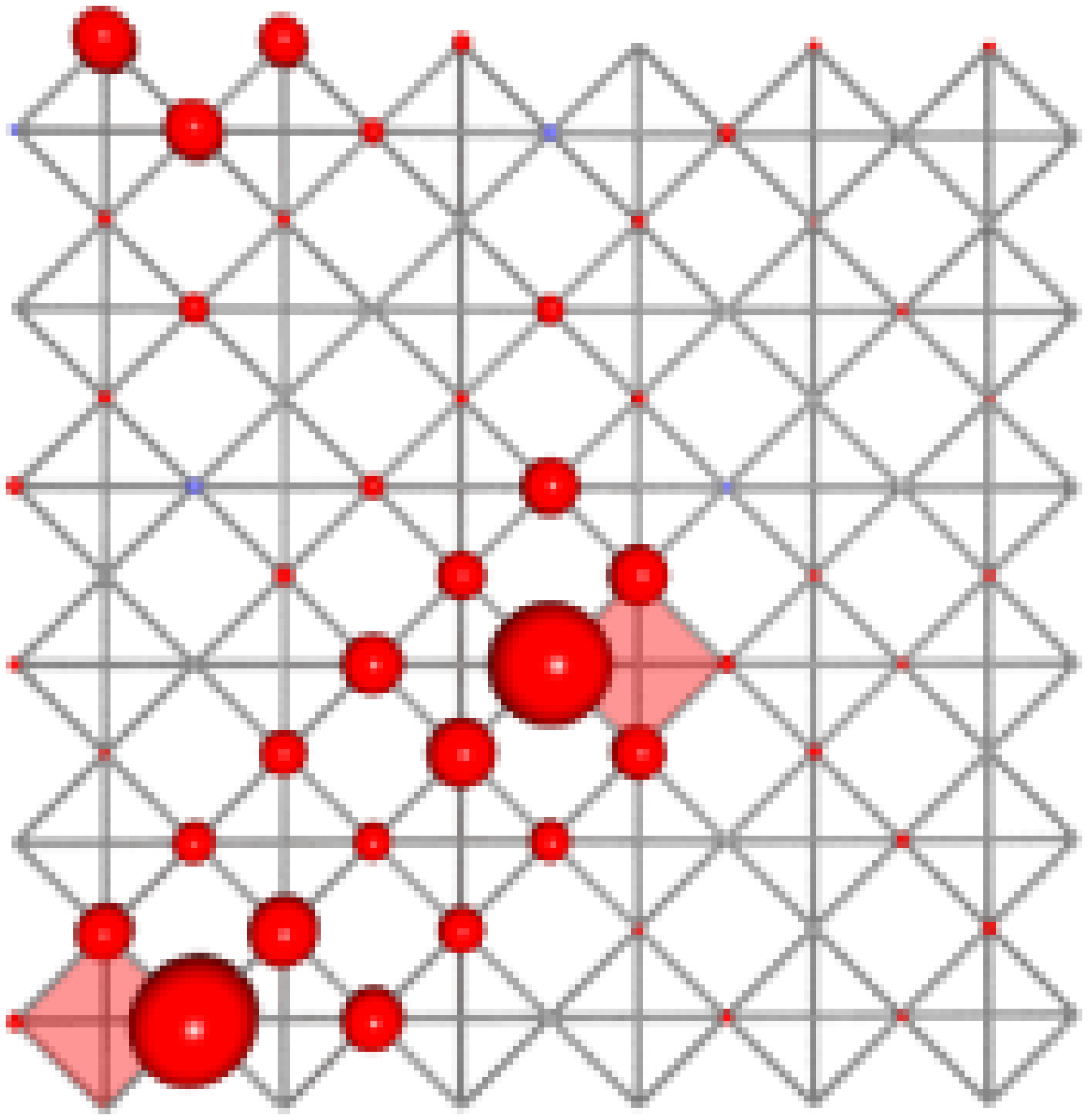}&
(e)\includegraphics[%
  width=27mm]{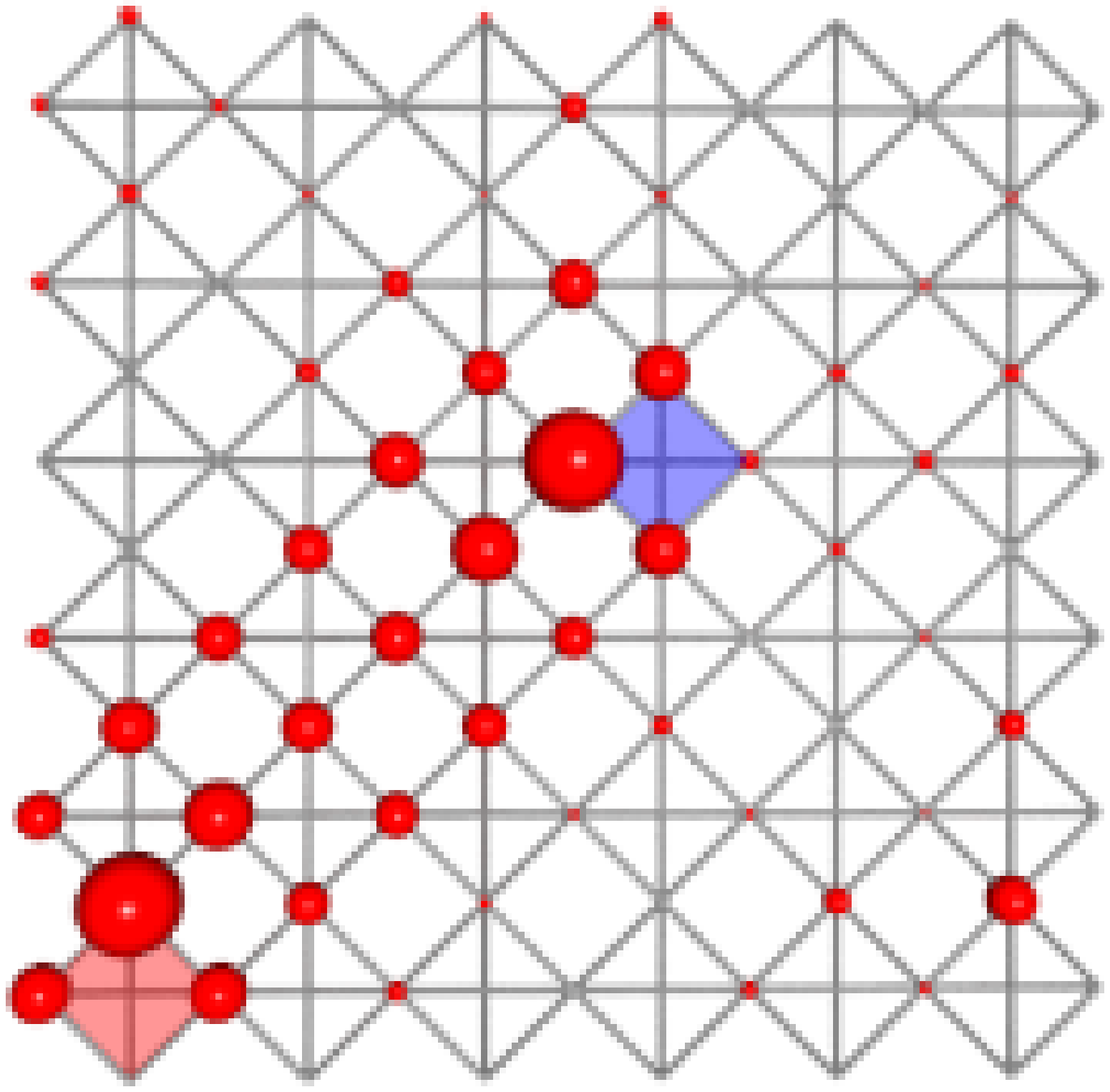}\\
(c)\includegraphics[%
  width=27mm]{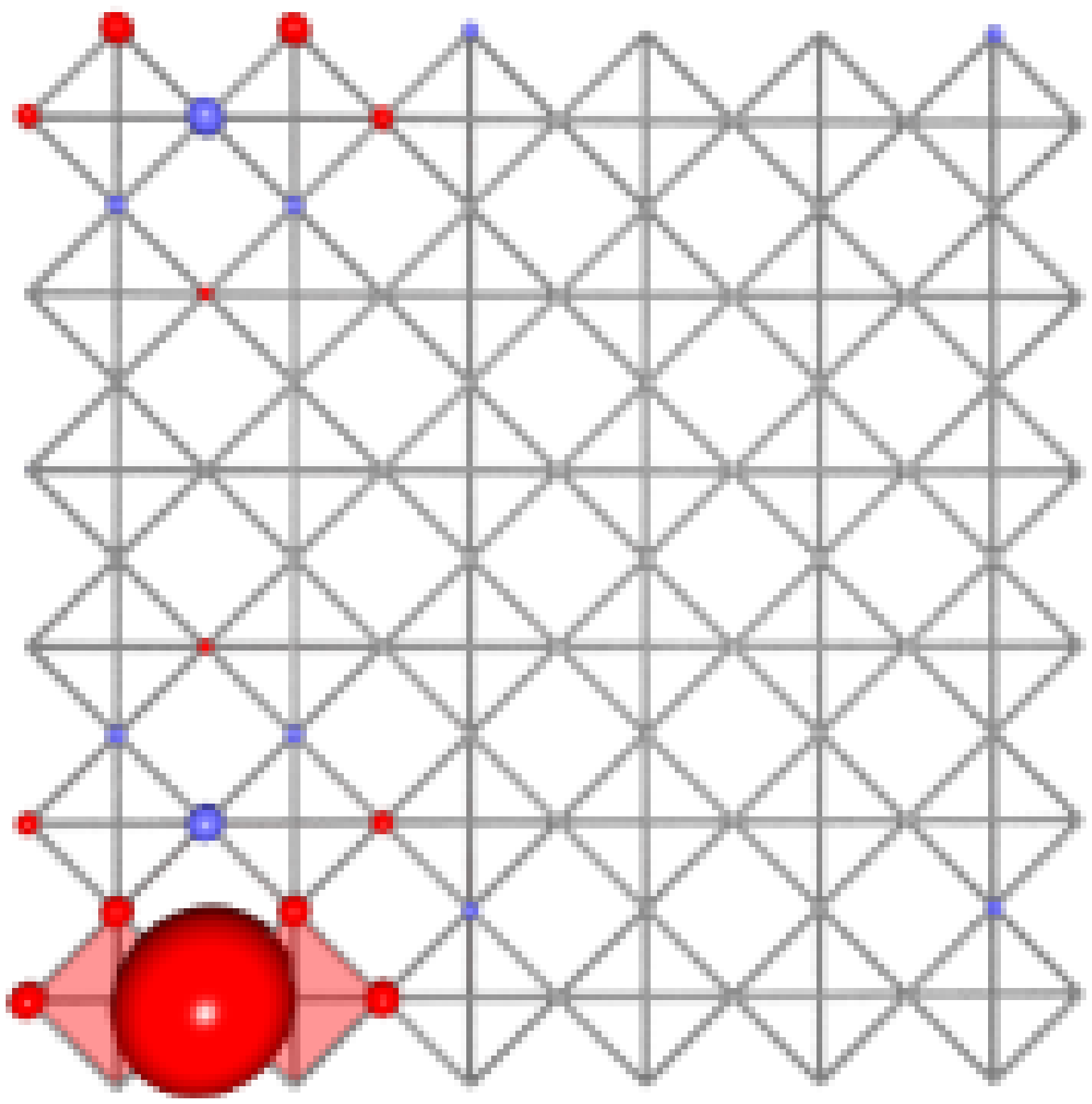}&
(d)\includegraphics[%
  width=27mm]{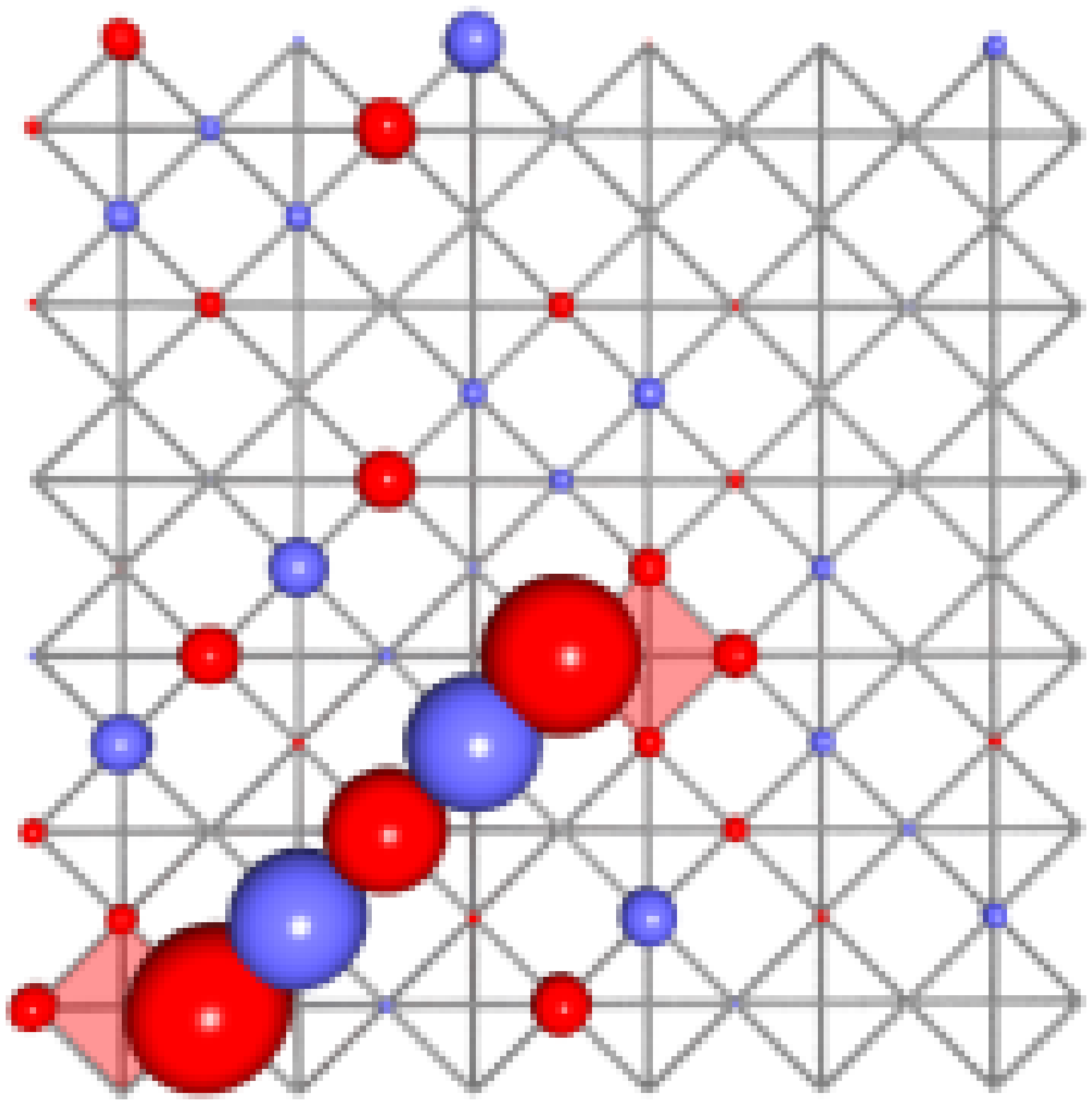}&
(f)\includegraphics[%
  width=27mm]{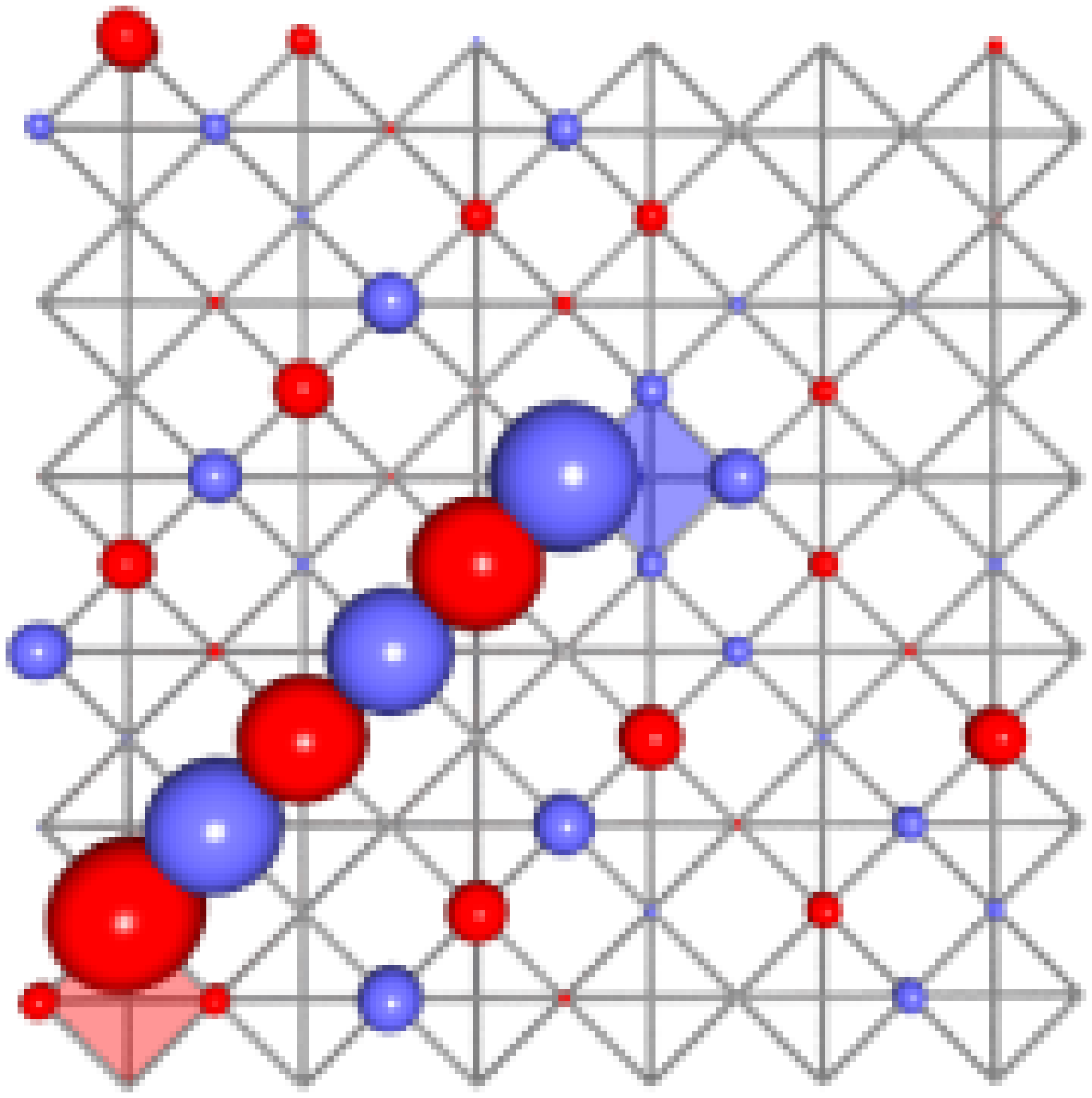}
\end{tabular}
\end{center}

\caption{(a), (b), and (e): Local loss of kinetic energy due to the separation
of two fractionally (static) charged defects (fcp's marked by light
red squares or fch's marked by dark blue squares). The radii of the
circles are proportional to the local energy loss. (c), (d), and (f):
Red (blue) circles show an increase (decrease) of the local density
(vacuum polarization due to the two fcp's or the fcp--fch pair).\label{cap:C4_string}
\label{cap:C4_string_fluct} }
\end{figure}

\section{Static and dynamical properties of the doped system\label{sec:Dynamical-properties}}

Next, we turn to the important issue of confinement/deconfinement
of defects generated either particle--hole excitations or by weak doping. Two different
kinds of calculations can be done: In the first case---referred to
as \emph{static} charges---the defects are fixed to some lattice positions
and ground--state calculations involving $H_{\mathrm{eff}}$ are performed
in the restricted subspace spanned by the configurations with exactly
those defects at given positions. The total (free) energy is recorded
as function of the defect positions. Its spatial derivative is interpreted
as attractive or repulsive force. In the second case---referred to
as \emph{dynamic} charges---the defects propagate according to the
Hamiltonian (\ref{eq:C4_hamil_tg}). The existence or non--existence
of bound states is interpreted as confinement or deconfinement. 

\paragraph{Static charges.} If an additional particle is added/doped
to a half-filled checkerboard lattice, at least two crisscrossed squares
are occupied by three particles. It is easy to see that, e.g., for
the $t$--$\mu$~Hamiltonian in the limit of large positive $\mu$ (frozen
configurations) the ground--state energy of two \emph{static} fractional
charges is independent of the distance between. Thus static charges
are not attracted by a force, which suggests that dynamic fcp's are
deconfined at zero temperature in that limit. 

Also, it is obvious that at the RK point the total energy of a system
with two static charges is independent of the distance between them.
Thus dynamic fcp's are expected to be deconfined. 

We will not discuss systems with a general values of $\mu$ further, but focus on the physical
case $\mu=0$ and discuss the changes of the kinetic energy density
in the presence of two static charges $e/2$.\cite{pollmann2006c} 
The energy change can be decomposed into local
contributions $\epsilon_{i}$ from all ring--exchange processes involving
a given site $i$. An increase in kinetic energy in the region between
the two fractional charges, i.e., along the connecting string, is
 found and illustrated in Fig.~\ref{cap:C4_string}(a,b). The total
energy increase is approximately proportional to the length of the
generated string and implies at large distances a constant confining
force. The changes of the local energy density goes along with density
changes, as illustrated in Fig.~\ref{cap:C4_string}(c,d). 
Similar results are found when a particle
is removed from the ground state or when a particle--hole excitation
is generated out of the ground state, see Fig.~\ref{cap:C4_string_fluct}(c,d,f). 

\noindent The calculations have been performed in a reduced Hilbert
space in which only two static defects are present. In a calculation
within the full Hilbert space, or a subspace allowing at least an additional
fcp--fch pair, the energy would \emph{not} increase linearly to infinity,
but the connecting string is expected to break by creating additional
 pairs of defects when this energetically favorable. For the relevant parameters,
e.g., $g=0.01t$ and $V=10t$, this occurs when two fcp's have separated
over 1000 lattice sites. This effect is well known for the case of
confined quarks where pair production (quark anti--quark pairs) occurs
before the quarks have been separated to an observable distance.

The attractive constant force acting between two fractional charges
in the confining phase results from a reduction of vacuum fluctuations
and a polarization of the vacuum in the vicinity of the connecting
strings. These findings suggest that a number of features known from
QCD are also expected to occur in a modified form in  solid--state physics. Conversely,
one would hope that by studying frustrated lattices or dimer models
one might be able to obtain better insight into certain aspects of
QCD. 

\paragraph{Dynamic charges.} 
We turn now to the dynamical properties
of fcp's, in particular spectral functions and optical conductivity
for the half--filled checkerboard lattice. Numerical studies of finite clusters are again the method of choice,\cite{pollmann2006d} because conventional approximation schemes such as mean--field theories
or Green's function decoupling schemes are unable to describe the
strong local correlations expressed by the tetrahedron rule.
Diagonalizations with up to 50 sites were performed within the minimal
Hilbert space spanned by the configurations with the smallest possible
number of violations of the tetrahedron rule {[}half--filled system:
all allowed configurations plus those with one fcp--fch pair; doped
case: 2 fcp's or 2 fch's{]}.

The spectral function $A(\mathbf{k},\omega)=A^{-}(\mathbf{k},\omega)+A^{+}(\mathbf{k},\omega)$
of an interacting many--particle system is the sum of the probability amplitudes for adding (+)
to the $N$--particle ground--state system $|\psi_{0}^{N}\rangle$ or removing (--) from it a particle with momentum
$\mathbf{k}$ and energy $\omega$ ($\hbar=1$) 
\begin{eqnarray}
A^{+}(\mathbf{k},\omega) & = & \lim_{\eta\rightarrow0^{+}}-\frac{1}{\pi}\mbox{Im}\langle\psi_{0}^{N}|c_{\mathbf{k}}\ \frac{1}{\omega+i\eta+E_{0}-H}\  c_{\mathbf{k}}^{\dag}|\psi_{0}^{N}\rangle\\
A^{-}(\mathbf{k},\omega) & = & \lim_{\eta\rightarrow0^{+}}-\frac{1}{\pi}\mbox{Im}\langle\psi_{0}^{N}|\  c_{\mathbf{k}}^{\dag}\ \frac{1}{\omega+i\eta-E_{0}+H}\  c_{\mathbf{k}}|\psi_{0}^{N}\rangle.\end{eqnarray}
We use operators $c_{\mathbf{k}}^{\dag}=\frac{1}{\sqrt{N}}\sum_{j}e^{i\mathbf{r}_{j}\mathbf{k}}c_{j}^{\dag}$
in the extended Brillouin zone. 

The spectral functions yield direct insight into the dynamics of a
many--body system, as seen, e.g., in angular--resolved photoemission
spectroscopy (ARPES). Expectation values of the form $G(z)=\langle\psi_{0}|A\ \,(z-H)^{-1}\  A^{\dag}|\psi_{0}\rangle$
can conveniently be calculated numerically by the Lanczos continued
fraction method \cite{gagliano1987} or kernel polynomial expansion
\cite{silver1996}. We found essentially identical results for both
algorithms.\cite{pollmannDissertation} 
However, the implementation of the Lanczos method turned
out to be slightly faster. Well converged results were obtained already
after several hundred iterations.

\begin{figure}
\begin{center}\begin{tabular}{cc}
(a)\includegraphics[%
  height=35mm,
  keepaspectratio]{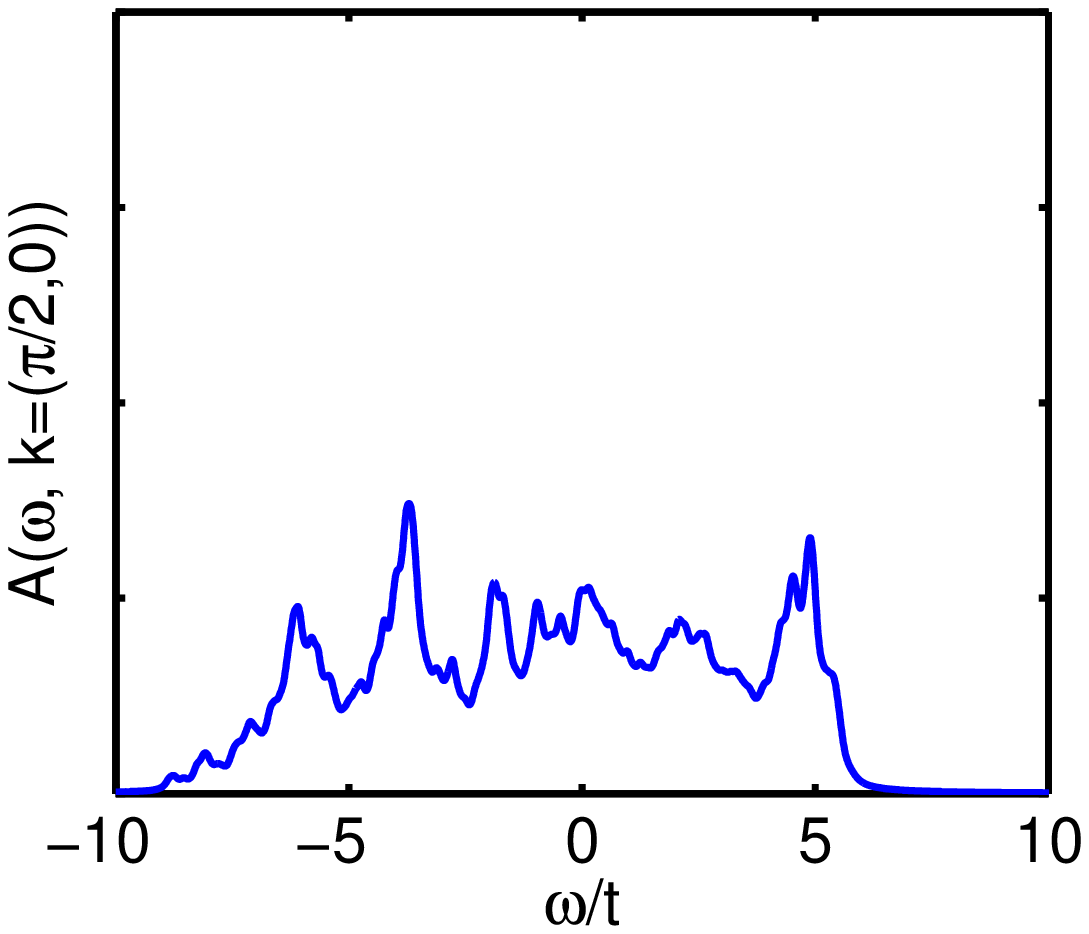}&
(b)\includegraphics[%
  height=35mm,
  keepaspectratio]{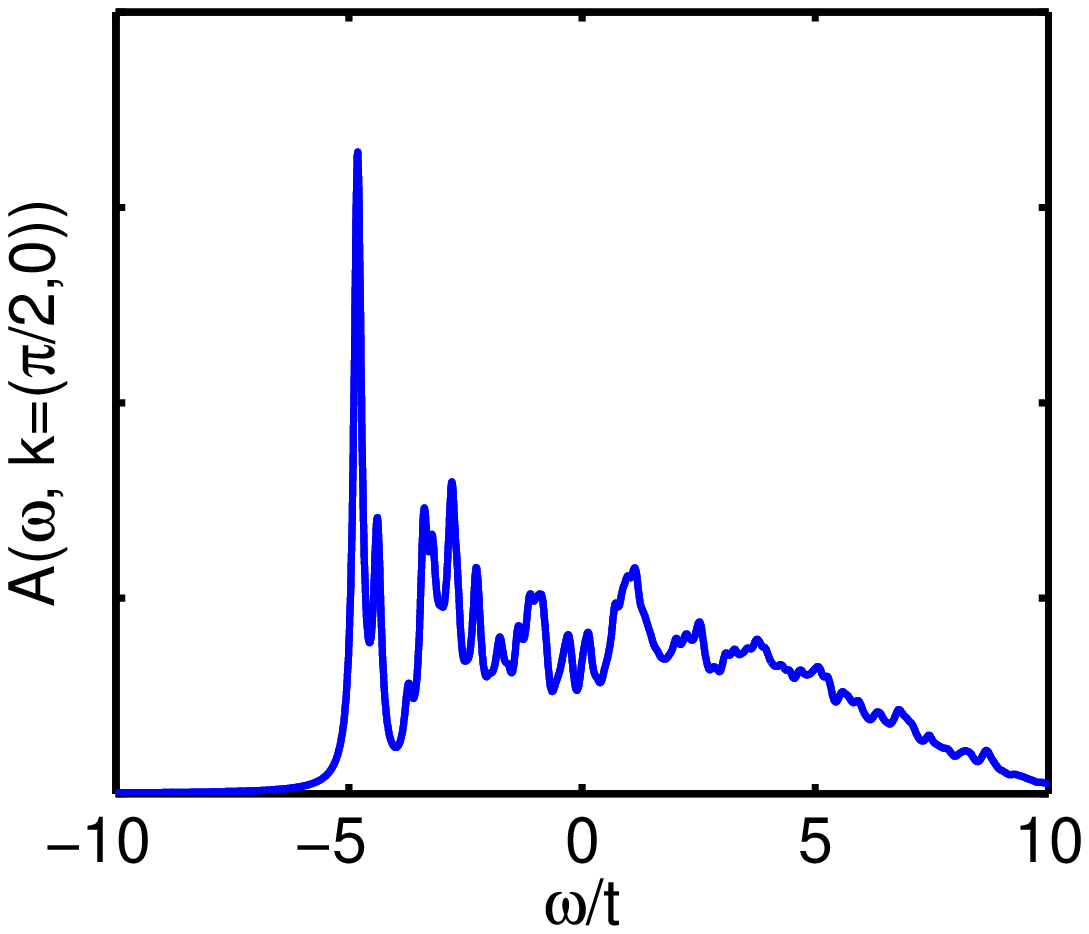}
\end{tabular}\end{center}

(a) Spectral function $A(\mathbf{k}=\pi/2,\omega)$ for $V=25t$ calculated
for a $\sqrt{32}\times\sqrt{32}$ cluster for the effective
$t$--$g$ Hamiltonian with (a) $g=0.01$ and (b) $g=1$. A Lorentzian broadening
$\eta=0.1t$ is used.\label{cap:Spectral-density_gg}
\end{figure}
 We checked numerically that in the limit of large $V$, it is sufficient
to calculate the spectral functions within the minimal Hilbert space. This
enables us to study rather large systems and to address the question
whether or not the two defects created by injecting one particle are
closely bound to one another or not. 

The concept of a spatial separation of fcp's can be confirmed by comparing results
within the minimal Hilbert space with those from an artificially restricted
calculation in which a particle is prevented from decaying into two defects with charge $e/2$ each. In our finite--cluster calculations, a broad low--energy continuum is seen in the spectral functions for the unrestricted case, which
is missing when the restriction is imposed.\cite{pollmann2006c} The
respective bandwidths are about $13t$ and $8t$. This suggests a
simple interpretation: The dynamics of two separated fcp's having
a bandwidth of $\approx6t$ each due to six nearest neighbors would
explain the calculated $13t$, while a confined added particle has a
much smaller bandwidth $8t$, which is close to $6t$.

Further insight can be obtained if the amplitude of the ring exchange
$g$=$12t^{3}/V^{2}$ in the effective Hamiltonian is considered an
independent parameter which is no longer restricted to the regime $g$$\ll$$t$
as enforced by $t$$\ll$$V$. Figure~\ref{cap:Spectral-density_gg} compares
results for the {}``physical'' regime corresponding to the previously
considered parameters $g\sim t^{3}/V^{2}=0.01t$ with those for $g=t$. In the
latter case, the broad continuum at the bottom of the spectral density
vanishes and a sharp $\delta$--peak evolves instead. The latter should
be viewed as a Landau quasi-particle peak. This suggests
the following interpretation: The ring exchange term leads for the
undoped case to charge order, which in finite--cluster calculations
is destroyed when a particle is added by the separation of the two fcp's. These are weakly bound
to each other. If $g$$\ll$$t$, the distance of the
two fcp's is larger than the system size considered and thus the excitations
seem to be deconfined. An artificially increased $g$ leads to much
stronger confinement and the diameter of the bounded pair becomes
smaller or comparable to the system size. The pair acts at low energies
as one entity. The electron as a (however strongly modified) ``particle''
is recovered and a finite weight of the quasi-particle peak results.
These findings suggest that for the physical regime $V/t\approx10$
quasi-particles with a spatial extent over more than hundred lattice
sites are formed. The huge spatial extent is expected to lead to interesting
effect. E.g., one may find with increasing doping concentration a
transition from a {}``confined'' phase to an {}``fcp
plasma'' phase with yet unknown properties when the average distance
of bound fcp--fcp pairs falls below the diameter of a single bound fcp
pair. 
Other open questions for future research include a thorough investigation 
of the 3D pyrochlore lattice. 
Even though the checkerboard lattice and pyrochlore lattice show many similarities, 
there exist certain differences between them, e.g., 
due to the higher spatial dimensionality.\cite{hermele2004, pollmann2006e}  
Here, differences of the two lattices arise in the U(1) gauge theory 
which describes the low energy excitations of the considered systems. 
The related compact electrodynamic is always confining in 2+1 dimensions 
while it allows for the existence of a deconfined phase in 3+1 dimension.\cite{polyakov1977} 
Systematic exact diagonalization studies as well as the application of Monte Carlo techniques 
will provide a deeper insight and we hope to confirm the expected existence of a deconfined phase in a 3D pyrochlore lattice. 
In addition, a natural extension of the model considered in this paper is the inclusion of spin. 
This leads to a more realistic model and could provide a better link to experiments. 

\end{document}